\documentclass[fleqn,usenatbib]{rasti}

\usepackage{newtxtext,newtxmath}
\usepackage[T1]{fontenc}

\DeclareRobustCommand{\VAN}[3]{#2}
\let\VANthebibliography\thebibliography
\def\thebibliography{\DeclareRobustCommand{\VAN}[3]{##3}\VANthebibliography}

\usepackage{graphicx}
\usepackage{amsmath}
\usepackage{booktabs}
\usepackage{tikz}
\usetikzlibrary{arrows.meta,fit,calc}

\graphicspath{{figures/}}

\newcommand{\modax}{\textsc{modax}}
\newcommand{\diffrax}{\textsc{diffrax}}
\newcommand{\diffequgpu}{\textsc{DiffEqGPU.jl}}
\newcommand{\torchdiffeq}{\textsc{torchdiffeq}}

\title{Accelerating massive ensembles of ordinary differential equations}

\author[Berry et al.]{
  Lawrence Berry$^{1,2}$\thanks{E-mail: ljb200@cam.ac.uk}, Will Handley$^{1,2}$, Oliver Hahn$^{3,4}$ and Nils Schöneberg$^{5}$
  \\
  $^{1}$Institute of Astronomy, University of Cambridge, Madingley Road, Cambridge, CB3 0HA, U.K.\\
  $^{2}$Kavli Institute for Cosmology, University of Cambridge, Madingley Road, Cambridge, CB3 0HA, U.K.\\
  $^{3}$Department of Astrophysics, University of Vienna,  Türkenschanzstraße 17, 1180 Vienna, Austria\\
  $^{4}$Department of Mathematics, University of Vienna, Oskar-Morgenstern-Platz 1, 1090 Vienna, Austria\\
  $^{5}$Universitätssternwarte, Ludwig-Maximilians-Universität München, Scheinerstraße 1, 81679 München, Germany
}

\date{Accepted XXX. Received YYY; in original form ZZZ}
\pubyear{\the\year{}}

\begin{document}
\label{firstpage}
\pagerange{\pageref{firstpage}--\pageref{lastpage}}
\maketitle

\begin{abstract}
  Solving large ensembles of small, independent ordinary differential equations (ODEs) is a common task in computational science and engineering, arising for example in Bayesian parameter estimation, Monte Carlo uncertainty quantification, and the integration of uncoupled physical systems such as the linearised Einstein--Boltzmann equations of cosmology. Modern graphics processing units (GPUs) with thousands of cores are well suited to such embarrassingly parallel workloads, but realising this potential requires careful attention to GPU-specific architectural constraints including single-instruction, multiple-thread execution, limited fast memory caches and higher single-precision compute throughput. We present \modax{}, a Python library of GPU-accelerated ODE solvers specifically optimised for solving large ensembles of low-dimensional, independent problems. Our solvers are compatible with the JAX ecosystem but are written in a low-level, thread-based programming paradigm better suited to ensembles of highly divergent trajectories than their pure-JAX counterparts. Benchmarks on the Lorenz, Robertson and van der Pol lattice systems compare \modax{} against \diffrax{}, \diffequgpu{} and \torchdiffeq{} across ensemble size, dimensionality, and trajectory divergence. Our explicit Tsit5 solver achieves 30$\times$ speed-up over \diffrax{} on Lorenz systems, and our implicit Rodas5P solver achieves over 700$\times$ speed-up on Robertson systems, demonstrating the superiority of linearly implicit Rosenbrock--Wanner methods on GPUs. Our implicit solver scales well to higher dimensions thanks to the use of sparsity-exploiting techniques and data structures. We conclude by applying \modax{} to three practical examples from the field of cosmology -- Bayesian parameter estimation from primordial abundances, Monte Carlo uncertainty quantification of the global 21\,cm signal, and the computation of uncoupled Fourier modes in the primordial power spectrum -- and demonstrate significant efficiency improvements in our ability to model physical phenomena described by low-dimensional (<200D) ODEs. The library and all benchmark code are available at \url{https://github.com/lawrenceberry/modax}.
\end{abstract}



\section{Introduction}
\label{sec:introduction}

\subsection{Background}
\label{sec:background}

An ordinary differential equation (ODE) describes the evolution of a state vector $\mathbf{y}(t) \in \mathbb{R}^D$ according to a known function $f$:
\begin{equation}
  \frac{\mathrm{d}\mathbf{y}}{\mathrm{d}t} = f(t, \mathbf{y};\, \mathbf{p}), \qquad \mathbf{y}(t_0) = \mathbf{y}_0,
  \label{eq:ode}
\end{equation}
where $\mathbf{p} \in \mathbb{R}^P$ denotes a set of parameters governing the dynamics. Together with an initial condition $\mathbf{y}_0$ at time $t_0$, these define an initial value problem (IVP). Numerical ODE solvers approximate the solution of this IVP by advancing the state through a sequence of discrete time steps using finite-difference or quadrature-based formulae. Such solvers underpin a vast range of scientific and engineering applications, from simulating the time evolution of physical systems to training neural differential equations \citep{Chen2018, Kidger2022}.

A particularly demanding class of problems requires solving not one but $N \gg 1$ such systems simultaneously, differing only in their initial conditions $\mathbf{y}_0^{(i)}$ or parameters $\mathbf{p}^{(i)}$. Problems of this type arise naturally across many scientific domains:

\begin{itemize}
  \item \textit{Bayesian parameter estimation and inverse problems.} Inferring the parameters $\mathbf{p}$ of a physical system from observational data frequently requires evaluating the likelihood over a large ensemble of candidate parameter vectors. Each evaluation involves solving equation~(\ref{eq:ode}) forward in time, so Markov chain Monte Carlo (MCMC) or nested sampling inference for ODE-governed systems may demand millions of solves.
  \item \textit{Monte Carlo uncertainty quantification.} Propagating uncertainty in initial conditions or model parameters through a dynamical system requires solving equation~(\ref{eq:ode}) for many draws from the input distribution. Modern ensemble weather and climate forecasting systems integrate the same atmospheric model from many slightly perturbed initial conditions to quantify forecast uncertainty \citep[e.g.][]{Lorenz1963}, and tracking the evolution of a debris cloud -- and assessing the risk of the runaway collisional cascade known as Kessler syndrome \citep{Kessler1978} -- requires propagating thousands of independent orbital trajectories, each governed by gravitational and drag force ODEs with slightly different initial states.
  \item \textit{Uncoupled ODE systems.} Many physical models described by partial differential equations (PDEs) can be decomposed into a large number of mutually independent low-dimensional ODEs, such as linear PDEs that decouple into ODEs in their diagonal basis. For example, the Fourier modes of the translation-invariant Einstein--Boltzmann equations, which describe the evolution of cosmic microwave background (CMB) anisotropies in the early universe \citep{MaBertschinger1995, Seljak1996}. Modern codes like \textsc{camb} \citep{Lewis2000} and \textsc{class} \citep{Lesgourgues2011} compute the CMB power spectrum by solving hundreds to thousands of independent ODEs at a time \citep{Howlett2012}.
\end{itemize}

In each case, the individual ODEs are small (state dimension $D \lesssim 200$) and mutually independent (the solution to trajectory $i$ does not depend on trajectory $j$), which make them a good fit for modern hardware accelerators like graphics processing units (GPUs). These offer a very large number of computational cores best suited to embarrassingly parallel workloads that do not require much memory or collaboration between cores. Top-of-the-range GPUs like the NVIDIA H100 contain tens of thousands of cores and provide up to 67 TFLOPS of double-precision floating-point performance \citep{NVIDIAHopper2022}, compared to hundreds of cores and only 1--2 TFLOPS for a comparably priced CPU \citep{AMDEPYC2022}. For a large ensemble of ODEs that can be parallelised across thousands of independent compute streams, this presents a potential order-of-magnitude improvement in both time-to-solution and cost per solve.

However, the prevailing model of GPU computing to date has been shaped by machine learning, where a single large neural network occupies much of the device memory, parallelism is expressed through large tensor operations, and automatic differentiation is a central requirement. For the scientific applications considered here, in which a single ODE is too small to effectively utilise a GPU on its own, the more important axis of parallelism is across a large ensemble of independent problems, each with a small state and modest memory requirements. Efficient ensemble evaluation is therefore our primary objective, with gradients providing a useful additional capability. The machine-learning origins of several existing GPU ODE libraries are reflected in their design: \diffrax{} \citep{Kidger2022} and \torchdiffeq{} \citep{Chen2018} were developed with neural differential equations in mind. \diffequgpu{} \citep{Utkarsh2023} does target ensembles of small ODEs directly, but in Julia, outside the Python and JAX ecosystem in which many scientific programs, such as the nested sampler used in Section~\ref{sec:bbn}, typically run.

\subsection{Existing GPU-accelerated ODE libraries}
\label{sec:existing}

\subsubsection{\diffequgpu{}}
\label{sec:diffeqgpu}

\diffequgpu{} \citep{Utkarsh2023} is a purpose-built Julia library for GPU-accelerated ensemble ODE solving. It targets precisely the same regime of batch-parallel ensembles of small ODEs as this paper, making it the most similar baseline tool and our primary point of comparison in the benchmarks of Section~\ref{sec:results}.

It exposes two distinct parallelisation strategies under the hood: the \texttt{EnsembleGPUKernel} backend, which compiles the entire adaptive solve loop into a single GPU kernel in which each trajectory takes its own independent time steps, and the \texttt{EnsembleGPUArray} backend, which combines $N$ small ODEs into a single large ODE and solves the combined system using joint time steps and GPU-accelerated array operations. The authors point out that the former approach is better suited to large ensembles of highly divergent trajectories as the joint time steps taken by the latter approach bottleneck computation on the slowest trajectory within each batch, and the array operations incur separate GPU kernel launches whose overhead can dominate the cost of the solves for low-dimensional problems.

\subsubsection{\diffrax{}}
\label{sec:diffrax}

\diffrax{} \citep{Kidger2022} is a comprehensive JAX-based library for numerical integration of differential equations, supporting ODEs, SDEs, and controlled differential equations. Built on JAX's \texttt{lax.while\_loop} primitive for adaptive time-stepping, it provides a wide range of solvers and first-class support for automatic differentiation, and was designed primarily to solve neural differential equations with relatively high numbers of parameters and expensive right-hand sides compared to the problems we target in this work.

Applying it to an ensemble of ODEs via \texttt{jax.vmap} produces a compiled batched solver whose trajectories step through time independently like \diffequgpu{}'s \texttt{EnsembleGPUKernel} backend, but whose underlying execution is via a single global loop of vectorised array operations over the batch like \texttt{EnsembleGPUArray}. Although individual trajectories are solved more quickly thanks to independent time-stepping, the loop itself must keep iterating until the slowest trajectory is solved, so it also suffers from bottlenecking on ensembles of highly divergent trajectories.

\subsubsection{\torchdiffeq{}}
\label{sec:torchdiffeq}

\torchdiffeq{} \citep{Chen2018} is a library of ODE solvers written in \textsc{PyTorch}, which like \diffrax{} was designed with neural differential equations in mind. It does not implement any adaptive implicit methods, but rather focuses on explicit Runge--Kutta integrators and its support for reverse-mode gradients via the continuous adjoint method.

It solves ensembles by making the state a $(D, N)$ tensor and treating it as one large ODE, similar to \diffequgpu{}'s \texttt{EnsembleGPUArray} backend. Trajectories advance in unison with a single step size dictated by whichever trajectory demands the smallest step. The adaptive step loop itself runs in Python on the host, dispatching stage evaluations to the GPU and making only limited use of kernel fusion to reduce CPU--GPU synchronisation overhead. Each step therefore carries a large fixed overhead that is negligible for large, expensive systems, but dominant for ensembles of small ODEs.

\subsubsection{\textsc{GRADSOLVE}}

While this manuscript was in preparation, two concurrent papers presented the JAX library \textsc{GRADSOLVE} for GPU ODE ensembles, describing its application to astrophysical problems \citep{SpurioMancini2026a} and its computation of reverse-mode gradients via the exact discrete adjoint \citep{SpurioMancini2026b}. We acknowledge these independent contributions.

\section{Methods}
\label{sec:methods}

We implement two GPU-accelerated automatically differentiable ODE solvers optimised for massive ensembles of low-dimensional divergent trajectories in a new Python package called \href{https://github.com/lawrenceberry/modax}{\modax{}}: the explicit Tsitouras 5(4) method for non-stiff systems and the fifth-order Rodas5P Rosenbrock--Wanner method for stiff systems. Section~\ref{sec:architecture} reviews the architectural nuances of GPUs that have constrained the design of these solvers, and Section~\ref{sec:implementation} describes their technical implementation in more detail.

\subsection{Architectural constraints}
\label{sec:architecture}

Realising the theoretical arithmetic throughput of a GPU for ensemble ODE solving requires careful consideration of the underlying architecture. The constraints imposed by GPU architectures motivate a number of design choices made later in Section~\ref{sec:implementation}.

\subsubsection{SIMT parallelism and thread divergence}
\label{sec:thread_divergence}

To achieve high arithmetic throughput within a limited transistor and power budget, GPU designers adopt an execution model fundamentally different from that of CPUs. First introduced with NVIDIA's \textsc{Tesla} GPU architecture \citep{Lindholm2008} and known as single-instruction, multiple-thread (SIMT) parallelism, threads are scheduled in groups of 32, known as a \emph{warp}, and the 32 lanes of a warp are constrained to execute the \emph{same instruction} in every clock cycle, though each lane operates on its own data held in registers. In practice, this means GPU threads cannot operate entirely independently; instead, they must work in concert with the other lanes of their warp. Fortunately, this architecture is ideally matched to ensemble ODE problems: $N$ independent trajectories naturally provide $N$ independent streams of identical arithmetic work.

However, SIMT execution introduces a critical hazard when code contains conditional branches: if some threads in a warp take one branch while others take another, both branches must be executed serially; threads not following the active branch sit idle. This phenomenon is known as \emph{thread divergence} and it can severely reduce utilisation when branches are unbalanced. Adaptive ODE solvers adjust their step size dynamically by evaluating an error estimate after each proposed step: the step is accepted when the error is within tolerance, or rejected and retried with a smaller step size. When an ensemble of trajectories is solved simultaneously within a single warp, divergence in the number of steps required by different trajectories can cause the warp to serialise over the slowest trajectory, with all other threads idle. Similarly, differences in the number of Newton iterations required for root finding in implicit solvers introduce another source of thread divergence. Minimising the potential for divergent control flow is therefore a key design criterion for GPU-optimised ODE solvers.

\subsubsection{Memory access patterns}
\label{sec:memory}

GPUs typically consist of memory tiers with very different bandwidths and latencies, ranging from larger but higher-latency off-chip global memory (DRAM) to smaller but lower-latency on-chip caches (SRAM) and registers that are private to each \emph{streaming multiprocessor} (SM), the unit of the GPU that schedules and executes warps \citep{NVIDIAHopper2022}. Warps are launched in \emph{thread blocks}, logical units that the hardware assigns to SMs, and each SM holds several resident blocks at once. To hide the latency of slow global memory accesses, GPUs interleave the execution of the warps resident on an SM, allowing data loading for one warp to be overlapped with computation for another. The ratio of resident active warps to the architectural maximum per SM is known as \emph{occupancy}. However, resident warps compete for the SM's limited registers and fast shared memory, so the more of either each warp reserves, the lower the occupancy.

Fortunately, low-dimensional ODE systems occupy little register and cache space, enabling high warp occupancy. Furthermore, because each ODE solve is independent, no inter-trajectory communication via slow global memory is required, and many of the temporary variables associated with the solver -- intermediate stage vectors, error estimates, step size state -- can reside in the registers and shared memory of the SM performing the work, avoiding the latency penalties of global memory access.

\subsubsection{Floating-point precision}
\label{sec:precision}

Scientific computing has traditionally relied on 64-bit double-precision (FP64) arithmetic, and modern x86 CPUs provide extensive hardware support for this type. GPUs, by contrast, typically offer substantially fewer FP64 cores than FP32 cores: half on the NVIDIA H100 and 1/64th on the gaming-oriented RTX 4090 \citep{NVIDIAAdaLovelace2022}. The substantially higher single-precision throughput motivates us to consider carefully how much double-precision arithmetic is strictly necessary for correctness of our ODE solvers and whether any steps of the algorithms can be carried out in single precision without compromising solution accuracy. Single-precision data also takes up less space in the fastest memory caches, potentially allowing more warps to be resident on each SM and improving occupancy.

\subsubsection{Kernel launch overhead}
\label{sec:kernel_overhead}

Each invocation of a GPU kernel -- a compiled device program -- incurs a fixed overhead as the driver schedules the kernel, transfers parameters to the device, and initialises execution contexts on the target SMs. This overhead typically amounts to several microseconds per launch and is negligible for kernels performing substantial arithmetic work. However, the right-hand side function $f$ of a small ODE is often computationally cheap, comprising only a handful of arithmetic operations. In such cases, the kernel launch overhead can dominate the cost of each function evaluation, and a solver that issues many separate kernel calls per step -- for example, one per Runge--Kutta stage or one per linear algebra routine -- will be severely penalised. Minimising the number of distinct kernel launches is therefore critical.

\subsection{Implementation}
\label{sec:implementation}

\begin{figure}
  \centering
  \includegraphics[width=\columnwidth]{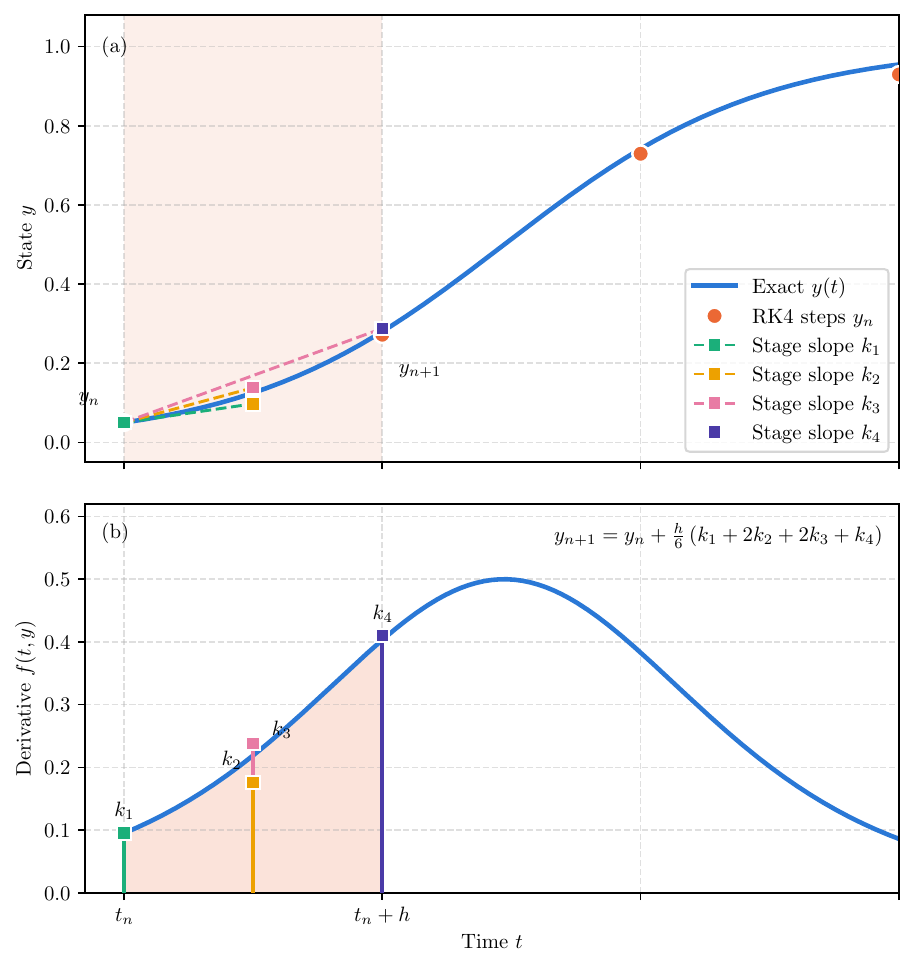}
  \caption{How a Runge--Kutta step integrates a trajectory, for the logistic equation $\dot y = 2y(1-y)$ solved by the classical fourth-order method with $h = 1$. (a) The exact solution and the sequence of accepted steps $y_n$. Within the highlighted step the right-hand side is evaluated at four stages. Each dashed line leaves $y_n$ with the slope $k_i$ of one stage and ends at the predicted state at which the next stage is evaluated -- an Euler step of $h/2$ along $k_1$ for $k_2$, of $h/2$ along $k_2$ for $k_3$, and of $h$ along $k_3$ for $k_4$. (b) The derivative $f$ along the trajectory. The exact increment $y_{n+1} - y_n$ is the shaded area under this curve, which the method approximates by the weighted sum of the stage slopes shown -- equivalent to Simpson's rule when $f$ is independent of $y$.}
  \label{fig:rk_quadrature}
\end{figure}

Integrating equation~\eqref{eq:ode} over one step from $t_n$ to $t_{n+1} = t_n + h$ gives the exact update:
\begin{equation}
  \mathbf{y}(t_{n+1}) = \mathbf{y}(t_n) + \int_{t_n}^{t_n + h} f\!\left(t, \mathbf{y}(t);\, \mathbf{p}\right) \mathrm{d}t,
  \label{eq:step_integral}
\end{equation}
so advancing the solution is a quadrature problem, with the complication that the integrand $f(t, \mathbf{y}(t))$ cannot be sampled at an interior point without already knowing the solution there. A Runge--Kutta method approximates the integral by a weighted sum of $s$ evaluations of $f$ at intermediate points, or \emph{stages}, within the step: each stage slope $\mathbf{k}_i = f(t_n + c_i h, \mathbf{y}_n + h\sum_{j} a_{ij}\mathbf{k}_j)$ is evaluated at a state that is itself predicted from the slopes already computed, and the step is completed as $\mathbf{y}_{n+1} = \mathbf{y}_n + h\sum_i b_i \mathbf{k}_i$ \citep{Hairer1993}. The coefficients $a_{ij}$, $b_i$ and $c_i$ form the method's Butcher tableau and are chosen so that the update matches the Taylor expansion of the true solution to a given order; the classical fourth-order method in Fig.~\ref{fig:rk_quadrature} reduces to Simpson's rule when $f$ depends on $t$ alone. In an explicit method each stage depends only on earlier stages, so a step is a fixed sequence of function evaluations. In an implicit method the stages depend on one another and each step requires the solution of a system of equations, which is the price of stability on stiff problems (Section~\ref{sec:rodas5}).

Rather than a single tableau, practical solvers embed a second set of weights $\hat{b}_i$ of lower order that shares the same stages, so that the difference $\hat{\mathbf{e}}_n = h\sum_i (b_i - \hat{b}_i)\mathbf{k}_i$ estimates the local error at no extra cost. Comparing this estimate against the user's tolerance decides whether the step is accepted and how large the next step should be, and it is this adaptive loop that both solvers in \modax{} execute for every trajectory. Adaptive step-size selection is governed by a proportional--integral--derivative (PID) controller \citep{Soderlind2002} of the form:
\begin{equation}
  h_{n+1} = h_n \, \min\!\left(\phi_{\max},\, \max\!\left(\phi_{\min},\, 0.9\, \varepsilon_n^{\,e_1}\, \varepsilon_{n-1}^{\,e_2}\, \varepsilon_{n-2}^{\,e_3}\right)\right),
  \label{eq:pid}
\end{equation}
where $\varepsilon_n$ is the scaled error norm of the current step, defined in equation~\eqref{eq:error_norm}, and $\varepsilon_{n-1}$, $\varepsilon_{n-2}$ are those of the two most recently accepted steps. The exponents are $e_1 = -(k_I + k_P + k_D)/q$, $e_2 = (k_P + 2k_D)/q$ and $e_3 = -k_D/q$, where $(k_P, k_I, k_D)$ are user-exposed gains and $q$ sets the strength of the controller's response ($q=5$ for Tsit5 and $q=6$ for Rodas5P). The default gains $(0, 1, 0)$ recover an elementary integral controller, but non-zero proportional and derivative gains tend to produce smoother step-size sequences, fewer rejected steps and more robust behaviour on stiff and mildly oscillatory problems -- all of which reduce the variance in step count across an ensemble and thereby mitigate thread divergence. The error history is advanced only on accepted steps, and the growth and shrink factors $\phi_{\max}$ and $\phi_{\min}$ bound the change in step size per step.

The error norm $\varepsilon_n$ that drives this controller is a weighted root-mean-square (RMS) norm over the scaled state components,
\begin{equation}
  \varepsilon_n = \sqrt{\frac{1}{D}\sum_{d=1}^{D} \left(\frac{w_d\, \hat{e}_{n,d}}{\mathrm{atol} + \mathrm{rtol}\,\max\!\left(|y_{n,d}|, |y_{n+1,d}|\right)}\right)^2},
  \label{eq:error_norm}
\end{equation}
where $\hat{e}_{n,d}$ is the embedded error estimate for component $d$ and $w_d$ is a user-exposed per-component error weight. The weights allow the user to control how much each state variable contributes to the norm; a weight of zero removes a component from step-size control entirely. This is particularly valuable in physical applications where the state vector mixes quantities of very different magnitudes and physical importance. Without weighting, the RMS norm can be dominated by components that are large but physically uninteresting, forcing unnecessarily small steps; conversely, it can under-resolve small but observationally important components. Exposing the weights lets the user prioritise accuracy in the components that actually matter.

Both \modax{} solvers implement dense output for the requested output times. Rather than forcing the adaptive step sequence to land exactly on every output time, an accepted step evaluates the method-specific continuous extension $\mathbf{y}(t_n + \theta h)$ for $0 \le \theta \le 1$ from the stage increments already in hand. This keeps the controller free to choose step sizes from the local error estimate alone while still returning solutions on user-specified grids, and it means the two solvers take the same steps whether one or one thousand output times are requested.

\subsubsection{Choice of numerical methods}

We have chosen to implement single-step methods as opposed to multistep methods like the backward differentiation formula (BDF) methods that are common in the CPU world (where they underpin codes such as SUNDIALS CVODE, MATLAB's \texttt{ode15s}, and SciPy's \texttt{BDF}/LSODA \citep{Hairer1991}). A BDF method fits a polynomial through the last few accepted steps, dynamically adapting both the step size and the polynomial order. Though it requires fewer stage evaluations, reading and writing the multistep history buffer is costly on a GPU and precisely the bandwidth pressure we seek to avoid in the memory-bound regime of massive ensembles (Section~\ref{sec:memory}). Additionally, BDF's dynamic order selection compounds the issue of thread divergence (Section~\ref{sec:thread_divergence}). The additional stage evaluations of a single-step method, by contrast, are fixed, predictable, and map cleanly onto SIMT lanes.

We have also chosen to implement higher-order methods where available: compared to lower-order alternatives of equivalent accuracy, higher-order methods tend to improve GPU utilisation. Though they require more stage evaluations per step, they can achieve the same accuracy in fewer steps. These additional stage evaluations are a fixed, predictable cost that map well to SIMT parallelism (the stages for all $N$ trajectories can be evaluated simultaneously), as opposed to additional steps which extend the adaptive while loop and increase the risk of thread divergence (Section~\ref{sec:thread_divergence}).

\subsubsection{Optimally parallelising an ensemble of ODEs}

So far we have only considered how to solve a single ODE. Optimally exposing parallelism across \textit{many} ODE trajectories depends on how similar those trajectories are to one another. When trajectories are expected to require similar numbers of adaptive steps -- as is the case for similar initial conditions or parameters -- the most straightforward approach is to vectorise a single-trajectory solver over the ensemble with \texttt{jax.vmap}, as our \diffrax{} baseline does. Each trajectory keeps its own step size and its own accept-or-reject decision, but the batched adaptive while loop advances all $N$ trajectories in lockstep, one iteration at a time. The $N$ evaluations of $f$ map directly to $N$ independent streams of identical arithmetic on the GPU, fully exploiting SIMT parallelism without any inter-thread communication.

However, the shared while loop means that the batch terminates only when the slowest trajectory has converged. When trajectories are highly heterogeneous -- differing substantially in the number of adaptive steps they require -- faster trajectories waste iterations waiting for slower trajectories. Achieving truly independent parallel execution requires a custom CUDA GPU kernel: a low-level device program that assigns one or more trajectories to each GPU warp and allows each to execute its own adaptive solve loop to completion without synchronising with other warps. This is analogous to what \cite{Utkarsh2023} call the `ensemble kernel' approach.

JAX and XLA cannot directly express this execution pattern because XLA requires a statically known computation graph: when a \texttt{lax.while\_loop} is batched via \texttt{vmap}, all trajectories in the batch are constrained to iterate in lockstep, and no mechanism exists for individual trajectories to terminate early and release their GPU resources. Truly independent per-trajectory loops, with different termination points, require thread-level control flow that is only expressible in a low-level GPU programming framework.

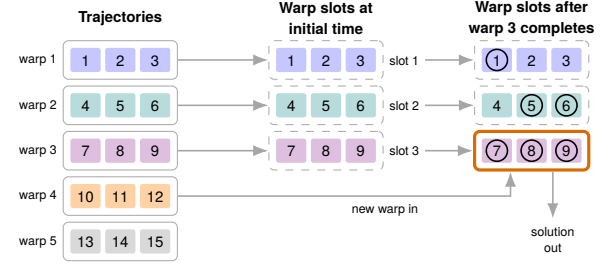
\begin{figure}
  \centering
\begin{tikzpicture}[
    x=1cm,y=1cm,
    >=Latex,
    font=\sffamily\scriptsize,
    flow/.style={-Latex, semithick, draw=gray!70},
    cell/.style={rounded corners=1pt, minimum width=.40cm,
                 minimum height=.32cm, inner sep=0pt},
    warprow/.style={draw=gray!65, rounded corners=2pt, inner sep=2.5pt},
    block/.style={draw=gray!65, dashed, rounded corners=2pt, inner sep=2.5pt},
    done/.style={circle, draw=black, semithick, inner sep=.4pt,
                 minimum size=.28cm},
    complete/.style={draw=orange!85!black, very thick, rounded corners=2pt,
                     inner sep=2.5pt},
    coltitle/.style={font=\sffamily\bfseries\scriptsize, align=center}
]
\colorlet{w1}{blue!22}
\colorlet{w2}{teal!28}
\colorlet{w3}{violet!25}
\colorlet{w4}{orange!35}
\colorlet{w5}{gray!30}
\def\dx{.46}   
\def\dy{.60}   

\node[coltitle] at (1.36,1.75) {Trajectories};
\foreach \r/\col in {0/w1,1/w2,2/w3,3/w4,4/w5} {
  \pgfmathsetmacro{\yy}{1.2-\dy*\r}
  \pgfmathtruncatemacro{\w}{\r+1}
  \node[anchor=east, font=\sffamily\tiny] at (.62,\yy) {warp \w};
  \foreach \c in {0,1,2} {
    \pgfmathtruncatemacro{\n}{3*\r+\c+1}
    \pgfmathsetmacro{\xx}{.90+\dx*\c}
    \node[cell, fill=\col] (tr-\r-\c) at (\xx,\yy) {\n};
  }
  \node[warprow, fit=(tr-\r-0)(tr-\r-2)] (warp-\r) {};
}

\node[coltitle] at (4.08,1.75) {Warp slots at\\initial time};
\foreach \r/\col in {0/w1,1/w2,2/w3} {
  \pgfmathsetmacro{\yy}{1.2-\dy*\r}
  \pgfmathtruncatemacro{\b}{\r+1}
  \foreach \c in {0,1,2} {
    \pgfmathtruncatemacro{\n}{3*\r+\c+1}
    \pgfmathsetmacro{\xx}{3.62+\dx*\c}
    \node[cell, fill=\col] (th0-\r-\c) at (\xx,\yy) {\n};
  }
  \node[block, fit=(th0-\r-0)(th0-\r-2)] (block0-\r) {};
  \node[anchor=west, font=\sffamily\tiny, inner xsep=2pt] (lab-\r) at (block0-\r.east) {slot \b};
  \draw[flow] (warp-\r.east) -- (block0-\r.west);
}

\node[coltitle] at (6.80,1.75) {Warp slots after\\warp 3 completes};
\foreach \r/\col in {0/w1,1/w2,2/w3} {
  \pgfmathsetmacro{\yy}{1.2-\dy*\r}
  \foreach \c in {0,1,2} {
    \pgfmathtruncatemacro{\n}{3*\r+\c+1}
    \pgfmathsetmacro{\xx}{6.34+\dx*\c}
    \node[cell, fill=\col] (th1-\r-\c) at (\xx,\yy) {\n};
  }
}
\foreach \r/\c in {0/0, 1/1, 1/2, 2/0, 2/1, 2/2} {
  \node[done] at (th1-\r-\c) {};
}
\node[block, fit=(th1-0-0)(th1-0-2)] (block1-0) {};
\node[block, fit=(th1-1-0)(th1-1-2)] (block1-1) {};
\node[complete, fit=(th1-2-0)(th1-2-2)] (block1-2) {};
\foreach \r in {0,1,2} {
  \draw[flow] (lab-\r.east) -- (block1-\r.west);
}

\draw[flow] ([xshift=.28cm]block1-2.south) -- ++(0,-.62)
  node[below, font=\sffamily\tiny, align=center] {solution\\out};
\draw[flow] (warp-3.east) -- ++(.25,0) -| ([xshift=-.28cm]block1-2.south)
  node[pos=.30, below, font=\sffamily\tiny] {new warp in};
\end{tikzpicture}
  \caption[Mapping ODE trajectories onto GPU warps and blocks]{How \modax{} maps an ensemble of ODE trajectories onto the GPU. Trajectories are queued in groups that each fill one warp, launched as its own thread block, and the hardware scheduler dispatches each block into a resident warp slot on a streaming multiprocessor (dashed boxes; an SM holds many such slots and interleaves the warps in them), where every thread integrates its own trajectory through an independent adaptive loop. Circled numbers mark trajectories whose solve has completed. A block retires only once every trajectory in its warp has completed, such as warp 3 here, at which point the solutions are read out and the next warp in the queue (warp 4) is slotted into the freed slot immediately, without waiting for the other warps to complete. Trajectories are therefore bottlenecked only on the slowest member of their own warp rather than on the slowest member of the entire ensemble. For clarity, the diagram shrinks warps from 32 threads down to 3 and shows only three of an SM's warp slots.}
  \label{fig:warp_scheduling}
\end{figure}

We therefore implement our solvers using \textsc{Numba-CUDA-MLIR}, which extends the LLVM-based \textsc{Numba} JIT compiler \citep{Lam2015} to CUDA device code, allowing us to write thread-level CUDA programs in plain Python. We assign one trajectory to each GPU thread in a 32-thread warp, with each warp executing its own adaptive solve loop to completion. Trajectories are therefore bottlenecked only on the slowest trajectory in each batch of 32, rather than across the entire ensemble. Each warp is launched as its own 32-thread block and occupies one of the resident warp slots of an SM, of which the Ada architecture used in Section~\ref{sec:results} provides 48 per SM, so many warps progress concurrently on each SM. Figure~\ref{fig:warp_scheduling} illustrates this scheduling: as soon as every trajectory in a warp has completed, its solutions are read out, the block retires, and the hardware scheduler slots the next block from the queue into the slot it freed. In addition, the resulting kernel is compiled into a single device program launched once at the start of each solve, eliminating the overhead of multiple kernel launches per step that \diffrax{} incurs. The compiled kernel is exposed to JAX as an XLA foreign function interface (FFI) custom call that allows these solvers to be called from within JAX programs without data copying or CPU--GPU synchronisation. A \modax{} \texttt{solve} is therefore an ordinary JAX primitive that can be traced under \texttt{jax.jit}. An outer \texttt{jax.vmap} over \texttt{solve} is intercepted and lowered to a single batched ensemble solve. Section~\ref{sec:sensitivity} describes how the \texttt{solve} primitive is also made differentiable.

\subsubsection{Tsit5: explicit solver}
\label{sec:tsit5}

For non-stiff systems we implement the Tsitouras 5(4) explicit Runge--Kutta pair \citep{Tsitouras2011}. The method is a seven-stage, fifth-order scheme with an embedded fourth-order error estimate. Its Butcher tableau satisfies only the first-column simplifying assumption, which frees enough coefficients to admit a fourth-order continuous extension that costs no additional function evaluations. The scheme is also \emph{first-same-as-last} (FSAL): the seventh stage of an accepted step is evaluated at the new solution point and is reused as the first stage of the next step, so each accepted step costs six rather than seven evaluations of $f$. Tsit5 is the recommended general-purpose non-stiff solver in the SciML and \diffrax{} ecosystems and is competitive with or superior to classical pairs such as Dormand--Prince 5(4) \citep{DormandPrince1980} on a wide class of problems. Its fixed seven-stage structure, with no inner loops other than the adaptive retry, is particularly well suited to SIMT execution: the stage work is identical for every thread in a warp, and the only source of divergence is the variable number of accepted and rejected steps across trajectories.

\begin{figure*}
  \centering
  \resizebox{\textwidth}{!}{
\begin{tikzpicture}[
    x=1cm,y=1cm,
    >=Latex,
    font=\sffamily\footnotesize,
    flow/.style={-Latex, semithick, draw=gray!70},
    lab/.style={font=\sffamily\footnotesize, fill=white, inner sep=1.5pt},
    box/.style={draw=gray!60, rounded corners=3pt, fill=gray!4,
    text width=3.2cm, align=center, inner sep=4pt},
    hub/.style={box, draw=orange!85!black, semithick, fill=orange!10},
    stg/.style={box, draw=violet!75!black, fill=violet!7},
    stgwide/.style={stg, text width=8.5cm},
    chk/.style={box, draw=teal!70!black, fill=teal!8},
    ttl/.style={font=\sffamily\bfseries\small},
    execute at begin picture={\def\ttl{\sffamily\bfseries\small}\thickmuskip=5mu\relax\medmuskip=4mu\relax}
  ]
  \def\xi{1.50}  \def\xl{5.30}  \def\xm{9.40}  \def\xr{14.30}
  \def\ya{0}     \def\yc{-2.35} \def\yd{-4.70} \def\ye{-6.40}

  \node[box, text width=2.8cm] (input) at (\xi,\ya)
  {{\ttl Input}\\[1pt] $\mathbf y_n \leftarrow \mathbf y_0$\\ $t_n \leftarrow t_0$, $h \leftarrow h_0$};
  \node[hub] (state) at (\xl,\ya)
  {{\ttl Current state}\\[1pt] $(\mathbf y_n,\ t_n,\ h)$\\ Loop while $t_n < t_f$};

  \node[stgwide] (fsal) at ({(\xm+\xr)/2},\ya)
  {{\ttl Stage 1 (FSAL)}\\[1pt] $\mathbf k_1 \leftarrow \mathbf k_7$ of the last accepted step,\\ else $\mathbf k_1 = f(t_n,\ \mathbf y_n)$};

  \node[stgwide] (stages) at ({(\xm+\xr)/2},\yc)
  {{\ttl Stages $i = 2,\ldots,7$}\\[2pt]
    \begin{tabular}{@{}l@{}}
      $\mathbf u_i = \mathbf y_n + h\sum_{j<i} a_{ij}\mathbf k_j$ (predicted state at $t_n + c_i h$)\\[1pt]
      $\mathbf k_i = f(t_n + c_i h,\ \mathbf u_i)$ (one $f$ evaluation)
  \end{tabular}};

  \node[chk] (check) at (\xr,\yd)
  {{\ttl Error check}\\[1pt] $\mathbf y_{n+1} = \mathbf y_n + h\sum_i b_i\,\mathbf k_i$\\
    $\hat{\mathbf e}_n = h\sum_i (b_i - \hat b_i)\,\mathbf k_i$\\
  Weighted norm $\|\hat{\mathbf e}_n\|_{\mathrm w} \le 1$?};
  \node[box] (pid) at (\xl,\yd)
  {{\ttl Step-size controller}\\[1pt] $h \leftarrow h \cdot \mathrm{factor}(\hat{\mathbf e}_n)$\\
  from the PID rule, eq.~\eqref{eq:pid}};

  \node[box] (acc) at (\xl,\ye)
  {{\ttl New state}\\[1pt] $\mathbf y_n \leftarrow \mathbf y_{n+1}$, $t_n \leftarrow t_n + h$\\
  $\mathbf k_1 \leftarrow \mathbf k_7$ (FSAL)};
  \node[box, text width=2.8cm] (out) at (\xi,\ye)
  {{\ttl Output}\\[1pt] Dense output at save\\ times in $[t_n, t_n + h]$};

  \draw[flow] (input) -- (state);
  \draw[flow] (state) -- (fsal);
  \draw[flow] (fsal.south) -- (stages.north);
  \draw[flow] (check.north |- stages.south) -- (check.north);
  \draw[flow] (check.south) |- (acc.east) node[lab, pos=.75, above] {Accept};
  \draw[flow] (check) -- node[lab, above] {Reject: discard $\mathbf y_{n+1}$} (pid);
  \draw[flow] (acc) -- (pid);
  \draw[flow] (acc) -- (out);
  \draw[flow] (pid.north) -- node[lab, left, align=center] {Next\\step} (state.south);
\end{tikzpicture}}
  \caption[The run-time loop of the Tsit5 kernel]{The run-time loop executed by each thread of the Tsit5 kernel, one adaptive step per iteration. There is no Jacobian, factorisation or inner iteration. The first stage derivative is inherited from the previous accepted step (FSAL) and evaluated afresh only at the start of the solve. The remaining six stages (violet) each form their own predicted state $\mathbf u_i$ from the earlier stage derivatives and evaluate $f$ once. The seventh stage sits at the end point $t_n + h$ and its predicted state is the fifth-order candidate $\mathbf y_{n+1}$ itself. The embedded fourth-order estimate $\hat{\mathbf e}_n$ (teal) is formed from the same seven derivatives at no extra cost. It is weighted according to equation~\eqref{eq:error_norm} and the step is either accepted -- advancing the state, carrying $\mathbf k_7$ over as the next $\mathbf k_1$ and writing dense output at any requested save times inside the step -- or rejected. In both cases the PID controller of equation~\eqref{eq:pid} rescales $h$ before the next step. The work per step is fixed and data independent, so the only source of divergence between the threads of a warp is the number of steps each trajectory takes.}
  \label{fig:tsit5_loop}
\end{figure*}
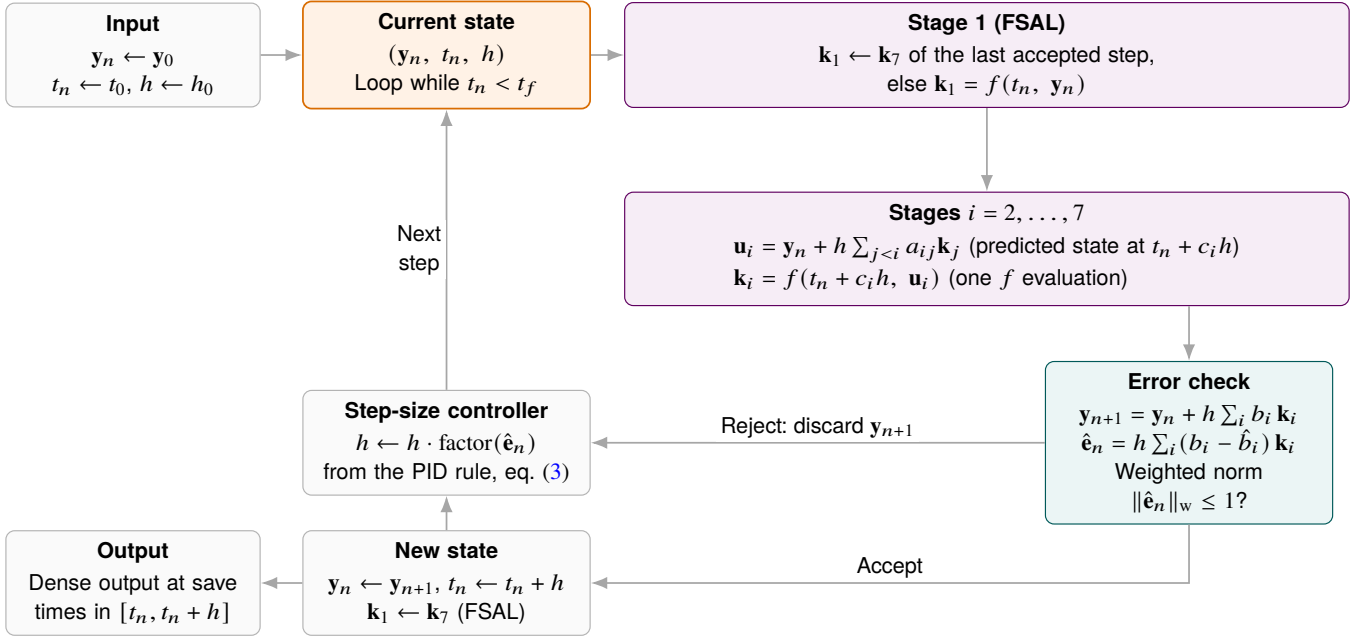

The kernel body is a direct transcription of the tableau presented in \citet{Tsitouras2011}. Every coefficient is a compile-time constant and every loop has a constant trip count, so each stage compiles down to straight-line arithmetic. Seven stage derivatives $\mathbf{k}_1, \ldots, \mathbf{k}_7$ are formed per step, the fifth-order candidate $\mathbf{y}_{n+1} = \mathbf{y}_n + h \sum_i b_i \mathbf{k}_i$ is proposed, and the embedded error estimate $\hat{\mathbf{e}}_n = h \sum_i (b_i - \hat{b}_i)\, \mathbf{k}_i$ computed. The error estimates are weighted and compared to tolerances according to equation~\eqref{eq:error_norm} to determine whether the step is accepted or not. If accepted, any requested output times falling inside the step are written out via dense interpolation as discussed, and $\mathbf{k}_7$ is carried over as the next $\mathbf{k}_1$. If rejected, the step is retried with a smaller step size prescribed by equation~\eqref{eq:pid}. Figure~\ref{fig:tsit5_loop} summarises this loop.

The nine per-trajectory working vectors -- the state, the trial state and the seven stage derivatives -- live either in fast on-chip shared memory or in thread-local memory, which resides in off-chip DRAM and is cached in L2 \citep{NVIDIACUDAGuide2024}. In the shared-memory kernel we lay out each vector as a $(D, 32)$ array so that consecutive 32-bit words are accessed by consecutive thread IDs, and an access is therefore fully coalesced when every lane of a warp touches the same relative address, such as the same index of an array variable \citep[Section~5.3.2, ``Device Memory Accesses: Local Memory'']{NVIDIACUDAGuide2024}. Shared memory costs $9 \times 32 \times D \times 8$ bytes per block, which caps the dimension at $D = 16$ on a typical consumer-grade GPU and limits occupancy. The solver therefore defaults to shared memory only when $D \le 16$ and $N \le 16\,384$ trajectories, and falls back to DRAM for larger systems and ensembles. As the GPU saturates at very large ensemble sizes, the latency of the accesses that reach the DRAM is hidden by occupancy.

\subsubsection{Rodas5P: stiff implicit solver}
\label{sec:rodas5}

For stiff systems, explicit methods are forced by stability rather than accuracy to take prohibitively small steps, and an implicit method is required. Standard implicit Runge--Kutta methods solve a non-linear system at every stage by Newton iteration, whose iteration count is data dependent and therefore a source of thread divergence (Section~\ref{sec:thread_divergence}). Rosenbrock methods \citep[Section~IV.7]{Hairer1991} avoid the iteration altogether by linearising each implicit stage equation about the current state with a first-order Taylor expansion. The result is a sequence of \emph{linear} systems, all sharing the same matrix, so each step performs one factorisation and a fixed number of triangular solves: a known, data-independent amount of work that maps cleanly onto SIMT lanes.

We implement the fifth-order, eight-stage Rodas5P method of \citet{Steinebach2023} in the transformed form of \citet[Section~IV.7]{Hairer1991}. With $J = \partial f / \partial \mathbf{y}$ and $f_t = \partial f / \partial t$ both evaluated at $(t_n, \mathbf{y}_n)$, the stage increments $\mathbf{k}_i$, $i = 1, \ldots, 8$, satisfy
\begin{equation}
  \begin{split}
    \left(\frac{1}{h\gamma} I - J\right) \mathbf{k}_i &= f\!\left(t_n + \alpha_i h,\ \mathbf{y}_n + \sum_{j<i} a_{ij}\mathbf{k}_j\right) \\
    &\quad + \frac{1}{h}\sum_{j<i} c_{ij}\mathbf{k}_j + d_i\, h\, f_t,
  \end{split}
  \label{eq:rosenbrock_stage}
\end{equation}
and the solution is advanced by
\begin{equation}
  \mathbf{y}_{n+1} = \mathbf{y}_n + \sum_{i=1}^{8} m_i \mathbf{k}_i,
  \label{eq:rosenbrock_update}
\end{equation}
where the coefficients $\gamma$, $a_{ij}$, $c_{ij}$, $\alpha_i$, $d_i$ and $m_i$ are transcribed from the Rodas5P tableau. The method is stiffly accurate, and the eighth stage is constructed so that $\mathbf{k}_8$ is itself the difference between the fifth-order solution and an embedded fourth-order one, so the error estimate $\hat{\mathbf{e}}_n = \mathbf{k}_8$ costs nothing extra. The last three stages share the end point $t_n + h$ and accumulate into the same argument vector. Rodas5P also supplies a continuous extension that is quartic in $\theta$ and built from the eight stage increments, which provides the dense output.

Rodas5P attains its full fifth order only when $J$ in equation~\eqref{eq:rosenbrock_stage} is the exact Jacobian. It is not a full W-method, whose order conditions would hold for an arbitrary $J$, but its coefficients do satisfy the additional W-method conditions up to second order, so an inexact $J$ reduces it to order 2, with order 1 for the embedded error estimator \citep{Steinebach2023}. The size of the additional error scales with the perturbation to $J$, and the step-size controller compensates for it by taking smaller steps. This robustness allows us to optionally carry out the factorisation and triangular solves in single precision for faster solves (Section~\ref{sec:precision}), where the perturbation is at the level of single-precision rounding.

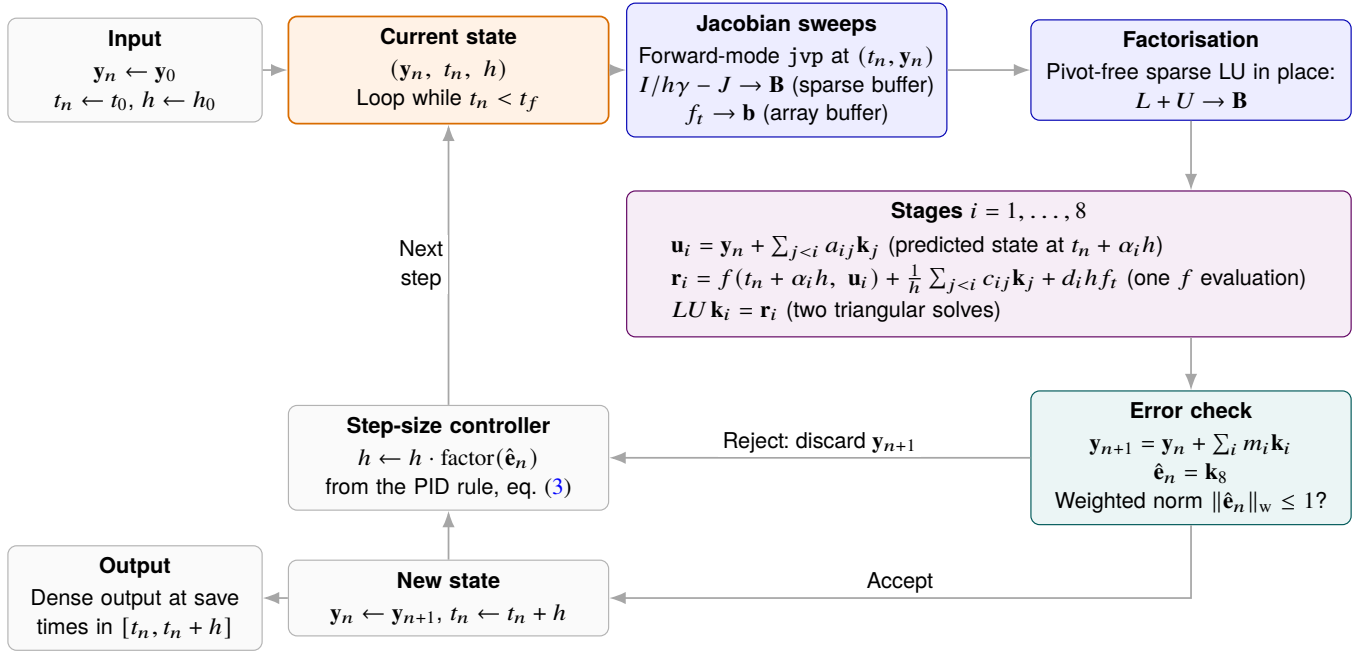
\begin{figure*}
  \centering
  \resizebox{\textwidth}{!}{
\begin{tikzpicture}[
    x=1cm,y=1cm,
    >=Latex,
    font=\sffamily\footnotesize,
    flow/.style={-Latex, semithick, draw=gray!70},
    lab/.style={font=\sffamily\footnotesize, fill=white, inner sep=1.5pt},
    box/.style={draw=gray!60, rounded corners=3pt, fill=gray!4,
    text width=3.6cm, align=center, inner sep=4pt},
    hub/.style={box, draw=orange!85!black, semithick, fill=orange!10},
    buf/.style={box, draw=blue!70!black, fill=blue!7},
    stg/.style={box, draw=violet!75!black, fill=violet!7,
    text width=8.5cm},
    chk/.style={box, draw=teal!70!black, fill=teal!8},
    ttl/.style={font=\sffamily\bfseries\small},
    execute at begin picture={\def\ttl{\sffamily\bfseries\small}\thickmuskip=5mu\relax\medmuskip=4mu\relax}
  ]
  \def\xi{1.50}  \def\xl{5.30}  \def\xm{9.40}  \def\xr{14.30}
  \def\ya{0}     \def\yc{-2.35} \def\yd{-4.70} \def\ye{-6.40}

  \node[box, text width=2.8cm] (input) at (\xi,\ya)
  {{\ttl Input}\\[1pt] $\mathbf y_n \leftarrow \mathbf y_0$\\ $t_n \leftarrow t_0$, $h \leftarrow h_0$};
  \node[hub] (state) at (\xl,\ya)
  {{\ttl Current state}\\[1pt] $(\mathbf y_n,\ t_n,\ h)$\\ Loop while $t_n < t_f$};

  \node[buf] (jac) at (\xm,\ya)
  {{\ttl Jacobian sweeps}\\[1pt] Forward-mode \texttt{jvp} at $(t_n,\mathbf y_n)$\\
  $I/h\gamma - J \rightarrow \mathbf B$ (sparse buffer)\\ $f_t \rightarrow \mathbf b$ (array buffer)};
  \node[buf] (lu) at (\xr,\ya)
  {{\ttl Factorisation}\\[1pt] Pivot-free sparse LU in place:\\
  $L + U \rightarrow \mathbf B$};

  \node[stg] (stages) at ({(\xm+\xr)/2},\yc)
  {{\ttl Stages $i = 1,\ldots,8$}\\[2pt]
    \begin{tabular}{@{}l@{}}
      $\mathbf u_i = \mathbf y_n + \sum_{j<i} a_{ij}\mathbf k_j$ (predicted state at $t_n + \alpha_i h$)\\[1pt]
      $\mathbf r_i = f(t_n + \alpha_i h,\ \mathbf u_i) + \frac{1}{h}\sum_{j<i} c_{ij}\mathbf k_j + d_i h f_t$ (one $f$ evaluation)\\[1pt]
      $L U\,\mathbf k_i = \mathbf r_i$ (two triangular solves)
  \end{tabular}};

  \node[chk] (check) at (\xr,\yd)
  {{\ttl Error check}\\[1pt] $\mathbf y_{n+1} = \mathbf y_n + \sum_i m_i \mathbf k_i$\\
  $\hat{\mathbf e}_n = \mathbf k_8$\\ Weighted norm $\|\hat{\mathbf e}_n\|_{\mathrm w} \le 1$?};
  \node[box] (pid) at (\xl,\yd)
  {{\ttl Step-size controller}\\[1pt] $h \leftarrow h \cdot \mathrm{factor}(\hat{\mathbf e}_n)$\\
  from the PID rule, eq.~\eqref{eq:pid}};

  \node[box] (acc) at (\xl,\ye)
  {{\ttl New state}\\[1pt] $\mathbf y_n \leftarrow \mathbf y_{n+1}$, $t_n \leftarrow t_n + h$};
  \node[box, text width=2.8cm] (out) at (\xi,\ye)
  {{\ttl Output}\\[1pt] Dense output at save\\ times in $[t_n, t_n + h]$};

  \draw[flow] (input) -- (state);
  \draw[flow] (state) -- (jac);
  \draw[flow] (jac) -- (lu);
  \draw[flow] (lu.south) -- (lu.south |- stages.north);
  \draw[flow] (check.north |- stages.south) -- (check.north);
  \draw[flow] (check.south) |- (acc.east) node[lab, pos=.75, above] {Accept};
  \draw[flow] (check) -- node[lab, above] {Reject: discard $\mathbf y_{n+1}$} (pid);
  \draw[flow] (acc) -- (pid);
  \draw[flow] (acc) -- (out);
  \draw[flow] (pid.north) -- node[lab, left, align=center] {Next\\step} (state.south);
\end{tikzpicture}}
  \caption[The run-time loop of the sparse Rodas5P kernel]{The run-time loop executed by each thread of the Rodas5P kernel, one adaptive step per iteration. With sparsity analysis disabled, the loop differs only in the matrix buffer, which becomes a row-major $D \times D$ array factorised with partial pivoting. Every step starts by rebuilding the iteration matrix in the thread's own compressed sparse row (CSR) buffer (blue): forward-mode \texttt{jvp} sweeps deposit $-J$ directly into the buffer and $f_t$ into a vector, $1/(h\gamma)$ is added to the diagonal, and the buffer is then overwritten in place by its pivot-free LU factorisation. The eight Rosenbrock stages (violet) each form their predicted state $\mathbf u_i$ from the earlier increments, evaluate $f$ once, assemble the right-hand side $\mathbf r_i$ of equation~\eqref{eq:rosenbrock_stage} and obtain $\mathbf k_i$ by a forward and a backward triangular solve against the same factorisation. The eighth stage doubles as the embedded error estimate (teal), which the weighted norm of equation~\eqref{eq:error_norm} either accepts -- advancing the state and writing dense output at any requested save times inside the step -- or rejects. In both cases the PID controller of equation~\eqref{eq:pid} rescales $h$ before the next step. The amount of work per step is fixed and data independent, so the only source of divergence between the threads of a warp is the number of steps each trajectory takes.}
  \label{fig:rodas5p_loop}
\end{figure*}

Like the Tsit5 solver, one trajectory is assigned to each thread in batches of 32, the minimum size of a warp. All working vectors are stored in local memory private to each thread. Each step (i) evaluates $-J$ and $f_t$ and writes $-J$ straight into the thread's matrix buffer, (ii) adds $1/(h\gamma)$ to the diagonal, (iii) factorises the result in place, (iv) runs the eight stages of equation~\eqref{eq:rosenbrock_stage}, (v) computes the error norm from $\mathbf{k}_8$ and (vi) applies the controller of equation~\eqref{eq:pid}, writing any output times inside an accepted step using the continuous extension. Figure~\ref{fig:rodas5p_loop} summarises this loop.

\paragraph{Computing the Jacobian.}
Implicit solvers require the Jacobian of the ODE. Since \textsc{Numba} compiles directly to LLVM and NVVM device code and is opaque to JAX, we instead use \textsc{Enzyme} \citep{Moses2020, Moses2021} to perform automatic differentiation on the compiled LLVM code itself. The \textsc{numba-enzyme} project \citep{NumbaEnzyme} wraps Enzyme for Numba on the CPU and we have developed a CUDA-capable fork, \textsc{numba-enzyme-cuda}, which differentiates CUDA device functions using primitives such as \texttt{jvp} and \texttt{vjp}. It emits derivatives as NVVM link-time-optimisation (LTO) IR code that the CUDA linker can inline into the solver kernel before the final optimisation pass. The primitives also compose with themselves to produce second derivatives (Section~\ref{sec:sensitivity}). This entire pipeline is invisible to the user, who supplies only $f$.

We use \emph{forward}-mode differentiation within \modax{}, and the directional-derivative Jacobian--vector product rather than the full Jacobian. In this case, where the dimensionality of outputs matches that of inputs, a forward sweep of the primal function $f$ is more efficient than a reverse sweep thanks to reduced memory requirements. A forward sweep seeded with a direction $\mathbf{v}$ over the flattened argument list $(y_1, \ldots, y_D, t, p_1, \ldots, p_P)$ returns $\partial f / \partial (\mathbf{y}, t, \mathbf{p}) \cdot \mathbf{v}$ at the cost of roughly one extra evaluation of $f$ in both compute and memory, so the full Jacobian $J$ and $f_t$ can be obtained at the cost of $D + 1$ primal evaluations. Sweeping one direction at a time keeps the working set at $\mathcal{O}(D)$ and allows us to fold each column into the matrix buffer as it is produced. Any zeroes in the seed vector are optimised away in the final compiler pass, as are any common subexpressions between sweeps.

\paragraph{Exploiting sparsity: colouring and ordering.}
The limited fast memory cache available in GPUs (Section~\ref{sec:memory}) motivates the use of sparse automatic differentiation techniques and sparse linear algebra routines to reduce the memory footprint of each ODE solve and improve occupancy, in addition to computational throughput. To this end, \modax{} has been built to exploit a Jacobian's sparsity pattern (which inputs affect which outputs) when it is already known at compile time.

Forward \texttt{jvp} sweeps are optimised using \emph{graph colouring} techniques. Columns of $J$ that do not share non-zero rows are structurally orthogonal: they do not contribute to the same index of $J\mathbf{v}$, so seeding both at once with a combined $\mathbf{v} = \mathbf{e}_i + \mathbf{e}_j$ returns both columns intact in a single sweep. Partitioning the columns into as few such groups as possible is a vertex colouring problem in which columns are adjacent if some row contains both \citep{Curtis1974, Coleman1983, Gebremedhin2005}. We run a heuristic greedy colouring algorithm from the \textsc{NetworkX} package \citep{Hagberg2008} to identify a good, if not optimal, partitioning scheme. The full Jacobian and $f_t$ then cost $n_{\mathrm{colours}} + 1$ sweeps instead of $D + 1$, which can be an order of magnitude fewer for a banded system.

Linear solves meanwhile are optimised by permuting the matrix to reduce fill-in and then factorising the permuted matrix in place with an optimal sparse data structure. At compile time, the variables are reordered to reduce fill-in using the implementation of the approximate minimum degree (AMD) algorithm of \citet{Amestoy1996} in the \textsc{SuiteSparse} package. The exact non-zero pattern of $L + U$ for the permuted matrix is obtained by symbolic Gaussian elimination, and then laid out in memory as a single compressed sparse row (CSR) matrix. The row pointers and column indices of this data structure are shared amongst all trajectories as either literals in the generated code or as a single table in constant memory depending on its size. Finally, we generate optimised factorisation and triangular solve functions over this compressed data structure. We avoid pivoting, which swaps rows by comparing magnitudes at run time, so that the non-zero pattern is fixed at compile time and identical in every thread. Since pivot-free elimination is not stable in general, we apply the fill-reducing permutation $P$ found by the AMD algorithm symmetrically, which preserves the diagonal entries of $M = I/(h\gamma) - J$. Thus the pivots are structurally present and contain $1/(h\gamma)$, which grows without bound as the step shrinks. Given Rodas5P remains stable to second order under an inexact Jacobian, a badly conditioned pivot turns into a matter for the step-size controller, whose response is a smaller step and hence a larger pivot.

\begin{figure*}
  \centering
  \resizebox{\textwidth}{!}{\input{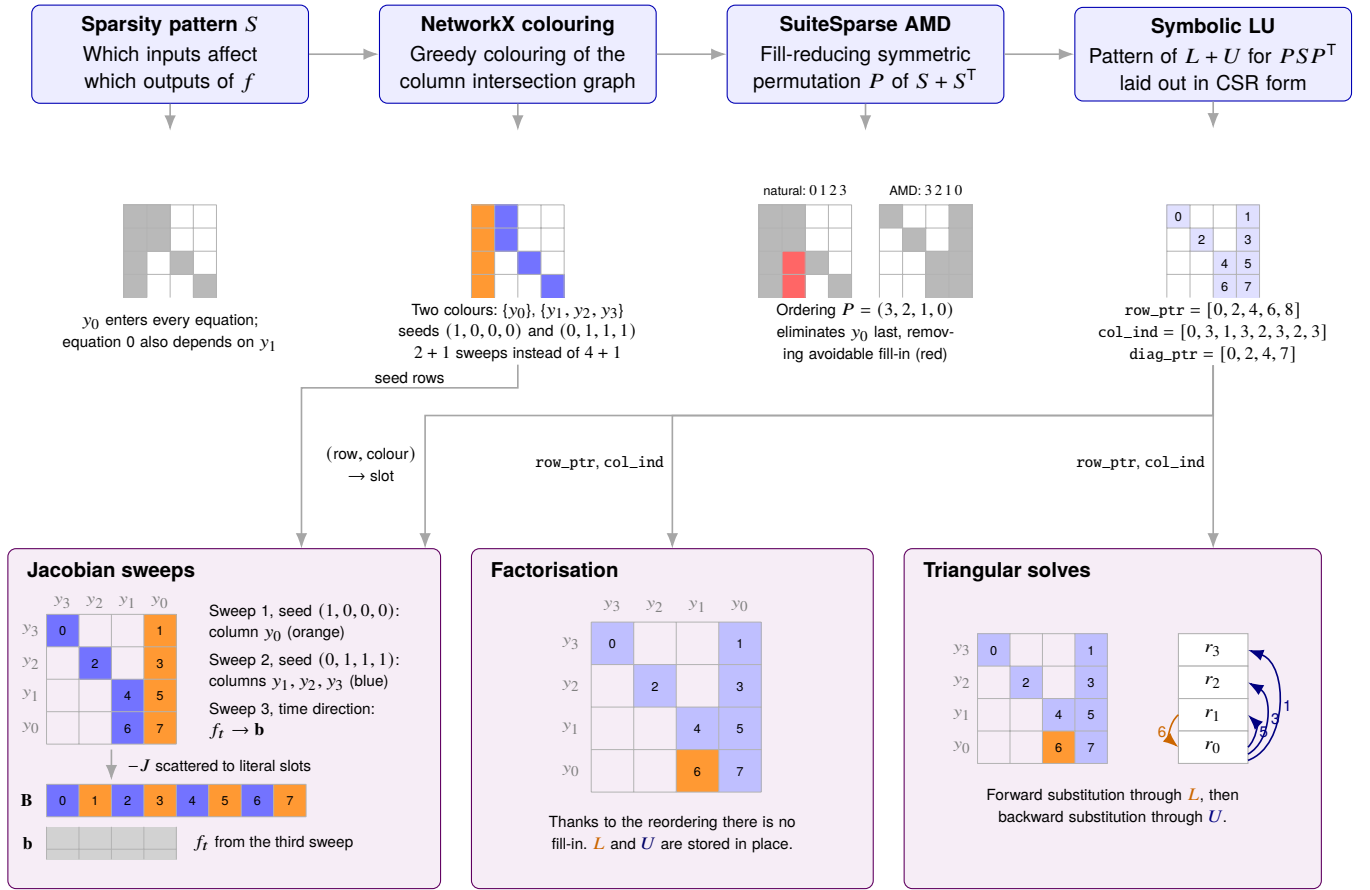}}
  \caption[Compile-time sparsity pipeline of the Rodas5P solver]{The compile-time sparsity pipeline of the Rodas5P solver, followed through a four-variable example. Top: the declared sparsity pattern $S$ is coloured by \textsc{NetworkX}, permuted by \textsc{SuiteSparse}'s approximate minimum degree algorithm and factorised symbolically. Middle: columns $y_1$, $y_2$ and $y_3$ share no non-zero rows and so take one colour (blue) while $y_0$ takes another (orange), which cuts the number of forward Jacobian sweeps required from $5$ to $3$. Eliminating $y_0$ first would fill two entries (red), whereas the AMD order $(3, 2, 1, 0)$ eliminates it last and fills none, so the CSR form of $L + U$ needs only 8 slots rather than 10. Bottom: the three compiled routines operate on the same $8$-slot buffer $\mathbf B$ with the number in each cell corresponding to its slot in that buffer. The Jacobian sweeps by colour are mapped into slots and deposited straight into the buffer. The factorisation writes $L + U$ in place. Thanks to the reordering there is no fill-in. The triangular solves consist of a row-by-row forward and then backward substitution pass through the buffer. All three routines are compiled into device code and linked into the Rodas5P kernel, with every slot index becoming a compile-time literal constant.}
  \label{fig:sparsity_pipeline}
\end{figure*}

To keep memory usage to a minimum, the optimised Jacobian sweeps write each (row, colour) pair of the compressed Jacobian directly into the CSR buffer, and the sparse factorisation then overwrites the buffer in place. The triangular solves read the buffer in elimination order but index the right-hand-side vector by the original variable numbers, so nothing is ever permuted at run time. When no sparsity pattern is supplied, each dimension of the system gets its own colour and the Jacobian matrix is dense and row-major, factorised by right-looking LU with partial pivoting as one would expect.
Fig.~\ref{fig:sparsity_pipeline} follows a four-variable example through this compile-time pipeline.

\subsubsection{Differentiation via forward sensitivity analysis}
\label{sec:sensitivity}

Our solvers support automatic differentiation of outputs via forward sensitivity analysis and are fully compatible with \textsc{JAX} automatic differentiation primitives such as \texttt{jax.jvp}. Considering the initial value problem posed in equation~\eqref{eq:ode}, provided $f$ is continuously differentiable, the sensitivity matrix $S(t) = \partial \mathbf{y}(t) / \partial \mathbf{p}$ obeys the variational equation:
\begin{equation}
  \frac{\mathrm{d}S}{\mathrm{d}t} = \frac{\partial}{\partial \mathbf{p}} f\!\left(t, \mathbf{y}(t; \mathbf{p}), \mathbf{p}\right) = J_y(t)\, S(t) + J_p(t), \qquad S(t_0) = 0,
  \label{eq:sensitivity}
\end{equation}
where $J_y = \partial f / \partial \mathbf{y}$ and $J_p = \partial f / \partial \mathbf{p}$, and the initial condition is zero because the initial state does not depend on the $P$ parameters. The sensitivity to the initial condition, $S_0(t) = \partial \mathbf{y}(t) / \partial \mathbf{y}_0$, obeys the same equation without the forcing term,
\begin{equation}
  \frac{\mathrm{d}S_0}{\mathrm{d}t} = J_y(t)\, S_0(t), \qquad S_0(t_0) = I,
  \label{eq:sensitivity_y0}
\end{equation}
since $\mathbf{y}_0$ enters $f$ only through the initial condition. Equations~\eqref{eq:sensitivity} and \eqref{eq:sensitivity_y0} constitute the classical forward, or direct, sensitivity method \citep{Dickinson1976, Feehery1997, Hindmarsh2005}. Each column of $S$ or $S_0$ is a $D$-vector obeying a linear ODE of the same form, and we write $S_k$ for the $k$-th such column and $J_{p,k}$ for its forcing term, which is column $k$ of $J_p$ for a parameter and zero for a component of $\mathbf{y}_0$. Stacking the $m$ columns actually being differentiated -- at most $P$ for the parameters plus $D$ for the initial state, as decided from JAX's symbolic zeros at trace time -- beneath the state gives one joint system in the augmented state $\mathbf{z} = (\mathbf{y}, S_1, \ldots, S_m)$:
\begin{equation}
  \frac{\mathrm{d}\mathbf{z}}{\mathrm{d}t} =
  \frac{\mathrm{d}}{\mathrm{d}t}
  \begin{bmatrix} \mathbf{y} \\ S_1 \\ \vdots \\ S_m
  \end{bmatrix}
  = F(\mathbf{z}) =
  \begin{bmatrix} f(t, \mathbf{y}, \mathbf{p}) \\ J_y S_1 + J_{p,1} \\ \vdots \\ J_y S_m + J_{p,m}
  \end{bmatrix}
  \label{eq:joint}
\end{equation}
of $D(1+m)$ components, which the existing kernels integrate jointly as a single ODE. Each sensitivity right-hand side $J_y S_k + J_{p,k}$ is $f$ differentiated in the direction $(S_k, 0, \mathbf{e}_k)$, or $(S_k, 0, 0)$ for an initial-state column, so it costs one forward sweep of the same Enzyme-differentiated $f$ that Rodas5P uses for its Jacobian, without ever forming $J_y$ or $J_p$.

For a non-stiff system, sensitivities can be obtained by integrating the joint system \eqref{eq:joint} with the usual explicit integrator. The function that evaluates the right-hand side emits $[f, J_y S_1 + J_{p,1}, \ldots, J_y S_m + J_{p,m}]$, the augmented state has $D(1+m)$ components in place of $D$, and the explicit integrator neither knows nor cares that the trailing components are sensitivities. By default the sensitivity components contribute to the error norm of equation~\eqref{eq:error_norm}, tying the accuracy of the gradient to the same tolerances as the values themselves.

For a stiff system the sensitivities cannot be handed to an explicit integrator, because they inherit the stiffness of the system. Since $f$ does not depend on $S$, the Jacobian of the joint system~\eqref{eq:joint} is block lower triangular,
\begin{equation}
  A = \frac{\partial F}{\partial \mathbf{z}} =
  \begin{bmatrix} J_y & 0 & \cdots & 0 \\ L_1 & J_y & & \\ \vdots & & \ddots & \\ L_m & & & J_y
  \end{bmatrix},
  \qquad
  L_k = \frac{\partial}{\partial \mathbf{y}}\left(J_y S_k + J_{p,k}\right),
  \label{eq:joint_jacobian}
\end{equation}
where the coupling block $L_k$ consists of second derivatives of $f$. Hence $\det(A - \lambda I) = \det(J_y - \lambda I)^{m+1}$ and the joint system has exactly the eigenvalues, and therefore stiffness, of the state equation and must be solved with an implicit integrator that requires $A$. A naive implementation would factorise this $D(1+m) \times D(1+m)$ matrix at a cost of $\mathcal{O}(D^3(1+m)^3)$ operations and $\mathcal{O}(D^2(1+m)^2)$ memory. Fortunately, the iteration matrix $I/(h\gamma) - A$ inherits the triangular structure of $A$ with the same block $M_0 = I/(h\gamma) - J_y$ on every diagonal, so each stage of equation~\eqref{eq:rosenbrock_stage} is a block forward substitution against a single factorisation of $M_0$,
\begin{equation}
  M_0\, \mathbf{k}_{y} = \mathbf{r}_{y}, \qquad M_0\, \mathbf{k}_{S_k} = \mathbf{r}_{S_k} + L_k\, \mathbf{k}_{y}, \quad k = 1, \ldots, m,
  \label{eq:block_forward}
\end{equation}
in which the state row is solved first and each sensitivity row is then solved against the same factors with its right-hand side corrected by the coupling term. Per step the cost is one factorisation of $M_0$ plus a pair of $\mathcal{O}(D^2)$ triangular solves per column per stage and the matrix buffer size remains unchanged. The overhead of computing gradients is therefore expected to grow linearly with respect to the number $m$ of gradients taken. Adjoint methods scale better with the number of gradients, but reverse-mode differentiation must revisit the forward trajectory in reverse order. The trajectory can be stored in full, which is impractical for large ensembles on memory-constrained GPUs; recomputed by integrating the state equation backward alongside the adjoint, which is unstable for stiff, dissipative systems \citep[Section~5.1.2]{Kidger2022}; or reconstructed from a small number of stored checkpoints between which forward segments are re-integrated, trading memory for extra solves \citep{Griewank2000}. Forward differentiation thus complements the primary objective of efficient ensemble solving without imposing impractical memory requirements or stability issues.

\section{Results}
\label{sec:results}

We evaluate \modax{} on three benchmark ODE systems that probe different aspects of solver performance: the Lorenz system (low-dimensional, non-stiff), the Robertson system (low-dimensional, stiff), and a coupled van der Pol lattice (medium-to-high-dimensional). All benchmarks were performed on a single NVIDIA GeForce RTX 4070 SUPER GPU, a consumer-grade device which has a similar number of CUDA cores and uses the same architecture as an L4-class datacentre GPU. For \diffequgpu{} benchmarks, we set \texttt{cpu\_offload = 0} to ensure all trajectories are solved on the GPU, ensuring a like-for-like comparison. Correctness of both \modax{} solvers has been verified separately against known analytic solutions, details of which are available in the test suite in the project repository.

\subsection{Lorenz system: non-stiff benchmarks}
\label{sec:results_lorenz}

The Lorenz system \citep{Lorenz1963}, originally derived as a model of convective flow, is the three-dimensional ODE
\begin{align}
  \dot{x} &= \sigma(y - x), \nonumber \\
  \dot{y} &= x(\rho - z) - y, \label{eq:lorenz} \\
  \dot{z} &= xy - \beta z, \nonumber
\end{align}
with $\sigma = 10$ and $\beta = 8/3$. With the classical value $\rho = 28$ the system is chaotic; our benchmarks instead use $\rho = 0.5$, below the pitchfork bifurcation at $\rho = 1$, where the origin is the unique globally stable fixed point and every trajectory decays exponentially towards it. We start trajectories far from the origin at $\mathbf{y}_0 = (1000, -500, 500)^\top$ and integrate over $t \in [0, 5]$ at $\mathrm{rtol} = 10^{-6}$ and $\mathrm{atol} = 10^{-8}$. The system is non-stiff, and in this regime the number of steps a trajectory requires is a monotone function of its initial distance from the origin, which gives direct control over the work each trajectory demands and makes it a convenient test bench for explicit solvers. We vary the ensemble size $N$ from $3$ to $10^5$ and the degree of divergence among trajectories from identical (all starting from the same initial condition) to heterogeneous (initial conditions drawn from a wide distribution that produces markedly different step counts).

\begin{figure}
  \centering
  \includegraphics[width=\columnwidth]{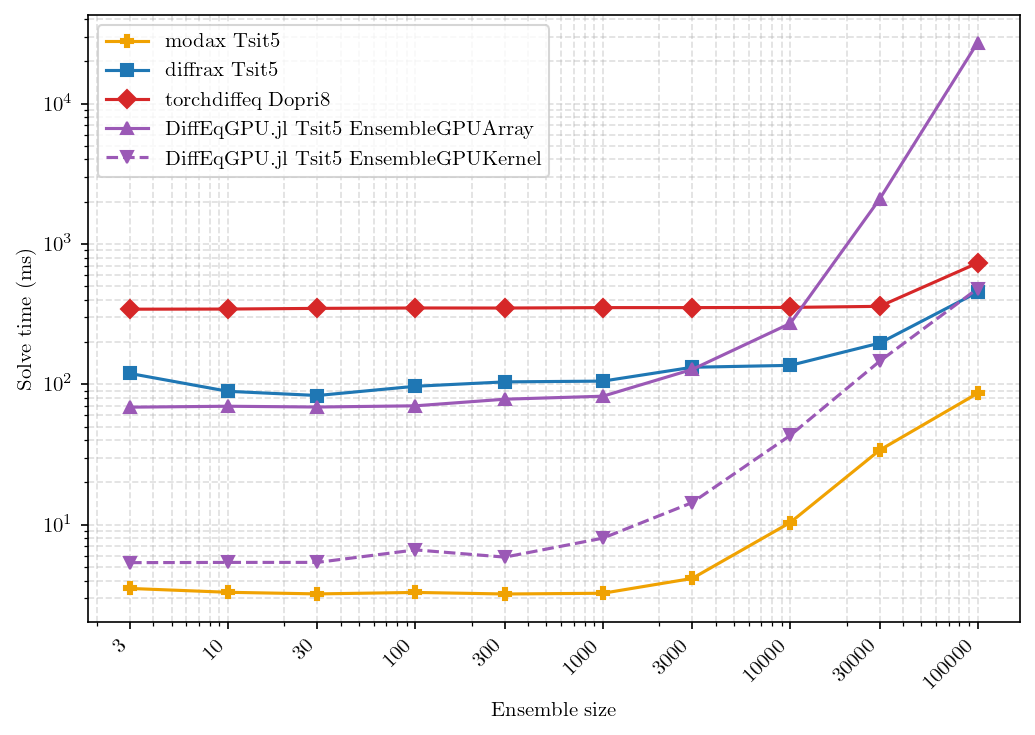}
  \caption{Wall-clock solve time as a function of ensemble size $N$ for identical Lorenz trajectories using the Tsit5 solver. \modax{} is roughly an order of magnitude faster than \diffrax{} across values of $N$.}
  \label{fig:lorenz_scaling}
\end{figure}

Figure~\ref{fig:lorenz_scaling} shows the wall-clock solve time of the four Tsit5 implementations as a function of ensemble size $N$ for identical trajectories. At $N=1000$ \modax{} achieves 30$\times$ speed-up over \diffrax{}, consistent with the elimination of essentially all kernel launch overhead. It is also 2.5$\times$ the speed of \diffequgpu{}'s EnsembleGPUKernel and scales slightly more favourably as $N$ increases. We speculate this is because our kernel has a lower register and cache footprint, which allows for higher warp occupancy: more trajectories are able to execute concurrently and hide memory latency before the scheduler must serialise them into successive waves.

\begin{figure}
  \centering
  \includegraphics[width=\columnwidth]{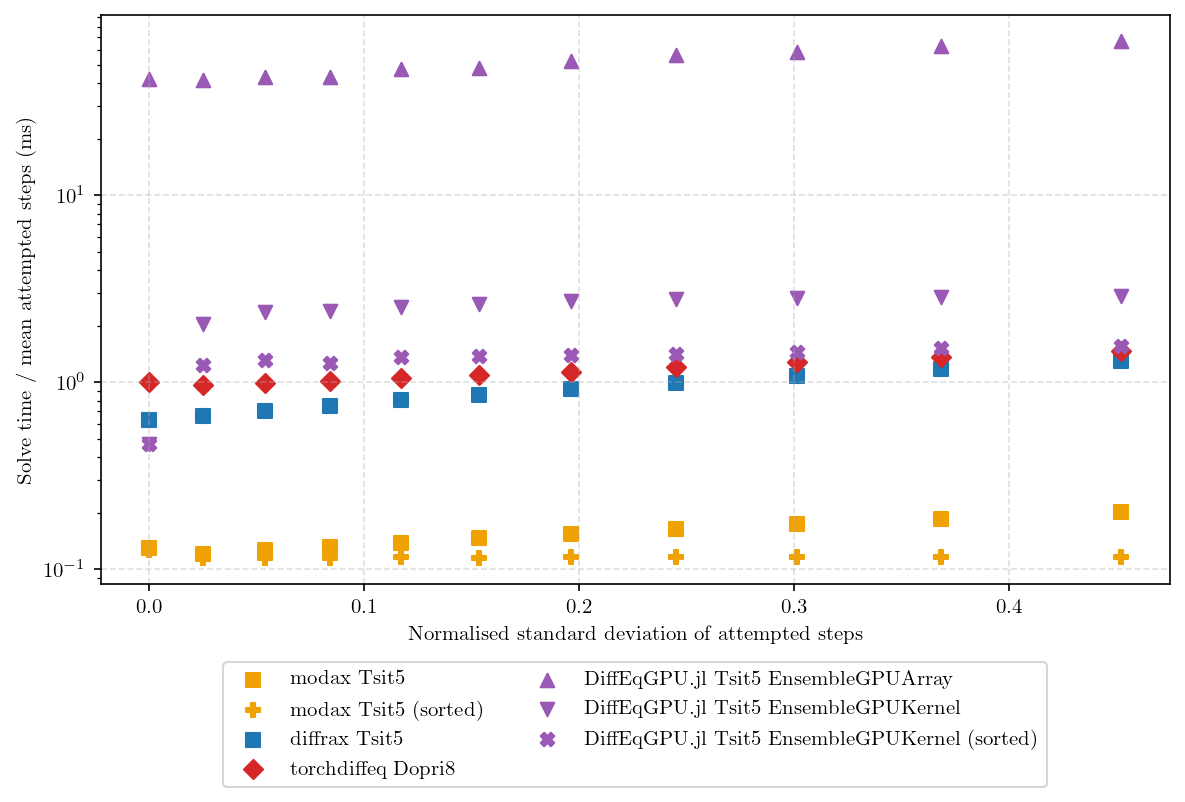}
  \caption{Sensitivity of explicit solvers to trajectory divergence for $N = 10^5$ Lorenz trajectories. Dopri8 is used for \torchdiffeq{} as it does not implement Tsit5. Normalised solve time is plotted against the standard deviation between trajectories of the number of attempted steps -- the greater the standard deviation, the more divergent the trajectories. In general, more divergent trajectories cause increased solve times, unless similar trajectories are batched together and solved independently.}
  \label{fig:lorenz_divergence}
\end{figure}

Figure~\ref{fig:lorenz_divergence} examines the sensitivity of the Tsit5 solvers to trajectory divergence at a fixed ensemble size of $N = 10^5$, chosen to saturate the GPU. Trajectory divergence is quantified by a parameter which controls the spread of initial distances to the origin. We report wall-clock solve times divided by the mean number of steps taken (to remove the trivial effect of harder trajectories requiring more steps overall) alongside the standard deviation, across the ensemble of trajectories, of the number of attempted steps, which measures how unevenly the computational burden is distributed across the ensemble. As expected, solve times increase with divergence, because harder trajectories force easier ones to iterate redundantly. \diffequgpu{}'s EnsembleGPUArray is more than an order of magnitude slower per step than every other solver throughout the sweep, consistent with its scaling in Fig.~\ref{fig:lorenz_scaling}.

When trajectories are pre-sorted by difficulty before being assigned to warps -- so that the most similar trajectories are co-located within the same warp -- \modax{}'s sensitivity to divergence is essentially eliminated. Trajectories that require similar numbers of steps are solved in the same warp, so no warp diverges significantly. By contrast, random assignment of divergent trajectories to warps can place the easiest and hardest trajectories in the same warp, maximising wasted computation. It is not always possible to trivially sort trajectories by difficulty, but sometimes heuristics derived from physical intuition can provide a good proxy.

\subsection{Robertson system: stiff benchmarks}
\label{sec:results_robertson}

The Robertson system \citep{Robertson1966} is a prototypical stiff chemical kinetics problem describing an autocatalytic reaction among three chemical species:
\begin{align}
  \dot{y}_1 &= -0.04\,y_1 + 10^4\,y_2\,y_3, \nonumber \\
  \dot{y}_2 &= \phantom{-}0.04\,y_1 - 10^4\,y_2\,y_3 - 3 \times 10^7\,y_2^2, \label{eq:robertson} \\
  \dot{y}_3 &= \phantom{-}3 \times 10^7\,y_2^2, \nonumber
\end{align}
The system is stiff due to the large disparity between the fast kinetic rates of the first two reactions and the slow evolution of the overall system. The classical initial condition $\mathbf{y}(0) = (1, 0, 0)^\top$ has a long induction period during which only the slow rate $0.04$ is active; our benchmarks instead start from $\mathbf{y}(0) = (0.891, 0.1, 0.009)^\top$, placing 10 per cent of the mass in the intermediate species $y_2$ so that the fastest reaction is active immediately and the solver must resolve the stiff timescale from its first step. Trajectories are integrated over $t \in [0, 10^5]$, with output at $t = 10^{-6}, 10^{-2}, 10^2$ and $10^5$, at $\mathrm{rtol} = 10^{-6}$ and $\mathrm{atol} = 10^{-8}$. It provides a demanding test of the Rodas5P implicit solver.

\begin{figure}
  \centering
  \includegraphics[width=\columnwidth]{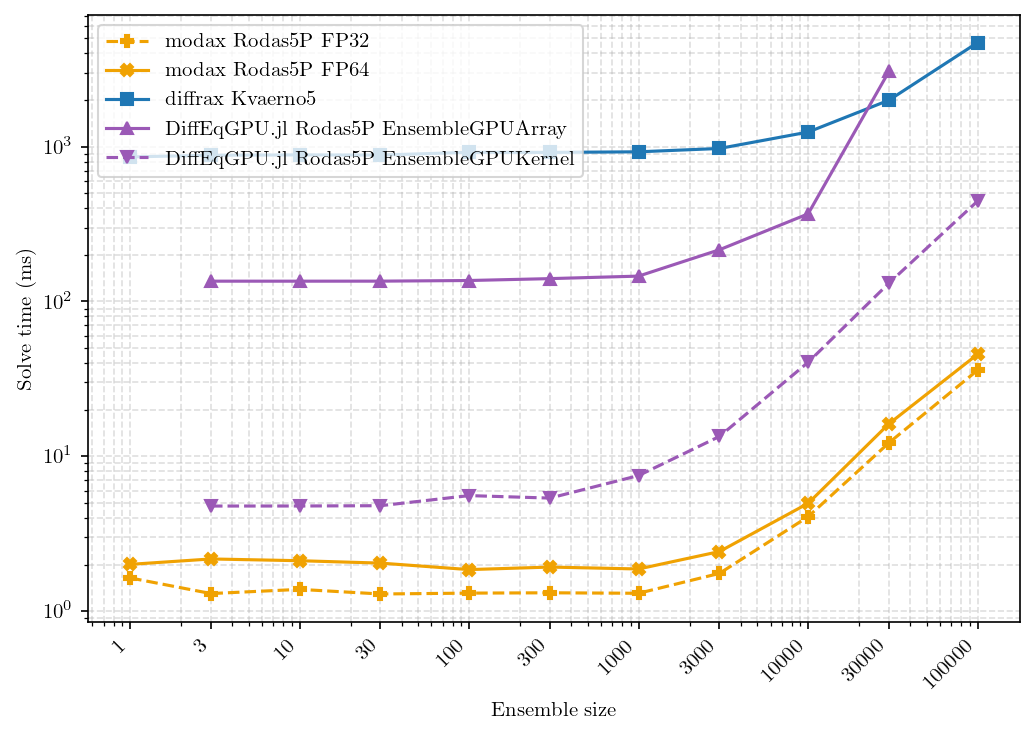}
  \caption{Wall-clock solve time as a function of ensemble size $N$ for identical stiff Robertson trajectories. \modax{}'s Rodas5P solver is roughly two to three orders of magnitude faster than the \diffrax{} Kvaerno5 solver. \diffequgpu{}'s $N=1$ point is omitted because it handles one-trajectory ensembles differently from batches, making that timing incomparable.}
  \label{fig:robertson_scaling}
\end{figure}

Figure~\ref{fig:robertson_scaling} shows how the Rodas5P solvers scale with ensemble size from $N = 1$ to $10^5$. Because \diffrax{} does not implement a Rodas5P solver, we include \diffrax{}'s fifth-order Kvaerno5 implicit Runge--Kutta method \citep{Kvaerno2004} as a reference. At $N=1000$ \modax{}'s Rodas5P solver is 6$\times$ faster than \diffequgpu{}'s EnsembleGPUKernel and 700$\times$ faster than \diffrax{}, which as a full implicit Runge--Kutta method requires iterative quasi-Newton root finding at each stage. The resulting variable iteration counts introduce severe thread divergence, illustrating the advantage of Rosenbrock methods' fixed-cost linear solves on GPUs.

\begin{figure}
  \centering
  \includegraphics[width=\columnwidth]{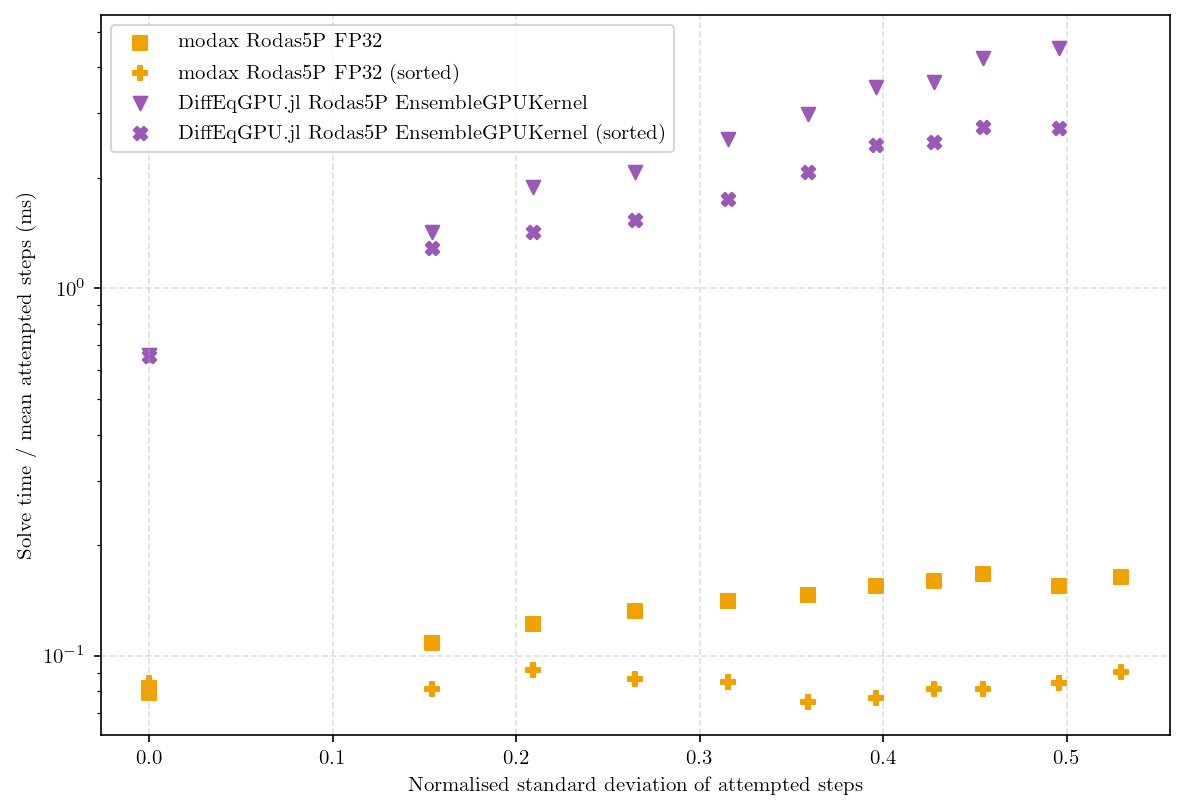}
  \caption{Sensitivity of implicit Rodas5P solvers to divergence for $N = 3 \times 10^4$ Robertson trajectories. Like in the Lorenz case, more divergent trajectories cause increased solve times, unless similar trajectories are batched together and these batches are solved independently.}
  \label{fig:robertson_divergence}
\end{figure}

Similarly to the Lorenz case, Fig.~\ref{fig:robertson_divergence} shows that increasing trajectory divergence substantially increases solve time for the ensemble of stiff Robertson trajectories. Here the divergence parameter $\delta \in [0, 3]$ draws the initial split of mass between $y_1$ and $y_3$ uniformly, moves the initial intermediate fraction $y_2(0)$ log-uniformly from $0.1$ towards $10^{-8}$, and scales each rate constant by an independent factor $10^{u}$ with $u \sim \mathcal{U}(-\delta, \delta)$, so that at $\delta = 3$ the rates vary by six orders of magnitude across the ensemble. However, when trajectories are pre-sorted by expected difficulty such that trajectories requiring a similar number of steps are grouped together in the same warp, \modax{}'s sensitivity to divergence is again essentially eliminated.

\subsection{Van der Pol lattice: scaling with dimensionality}
\label{sec:results_vdp}

So far we have considered only low-dimensional ODE systems ($D = 3$). The van der Pol oscillator \citep{VanDerPol1926} provides a natural vehicle for studying scaling with dimensionality. A single van der Pol oscillator evolves as
\begin{align}
  \dot{y}_1 &= y_2, \nonumber \\
  \dot{y}_2 &= \mu(1 - y_1^2)\,y_2 - y_1, \label{eq:vdp}
\end{align}
where $\mu > 0$ controls the degree of stiffness: for $\mu \gg 1$ the system is stiff, for $\mu = 1$ it is mildly non-stiff. To obtain a $D$-dimensional system we couple $D/2$ oscillators diffusively in a ring lattice. Writing $x_i$ and $v_i$ for the position and velocity of oscillator $i \in \{1, \dots, D/2\}$, the lattice evolves as
\begin{align}
  \dot{x}_i &= v_i, \nonumber \\
  \dot{v}_i &= \mu(1 - x_i^2)\,v_i - x_i + d\,(x_{i-1} + x_{i+1} - 2x_i), \label{eq:vdp_lattice}
\end{align}
with indices taken modulo $D/2$ so that the first and last oscillators are neighbours, and coupling strength $d = 10$. Its Jacobian has a banded (cyclic tridiagonal-block) sparsity pattern that couples every oscillator to the rest of the lattice whilst leaving the stiffness tunable via $\mu$. We use the non-stiff lattice ($\mu = 1$) to benchmark Tsit5 and the stiff lattice ($\mu = 100$) to benchmark Rodas5P, in both cases solving $N = 1000$ identical trajectories.

\begin{figure}
  \centering
  \includegraphics[width=\columnwidth]{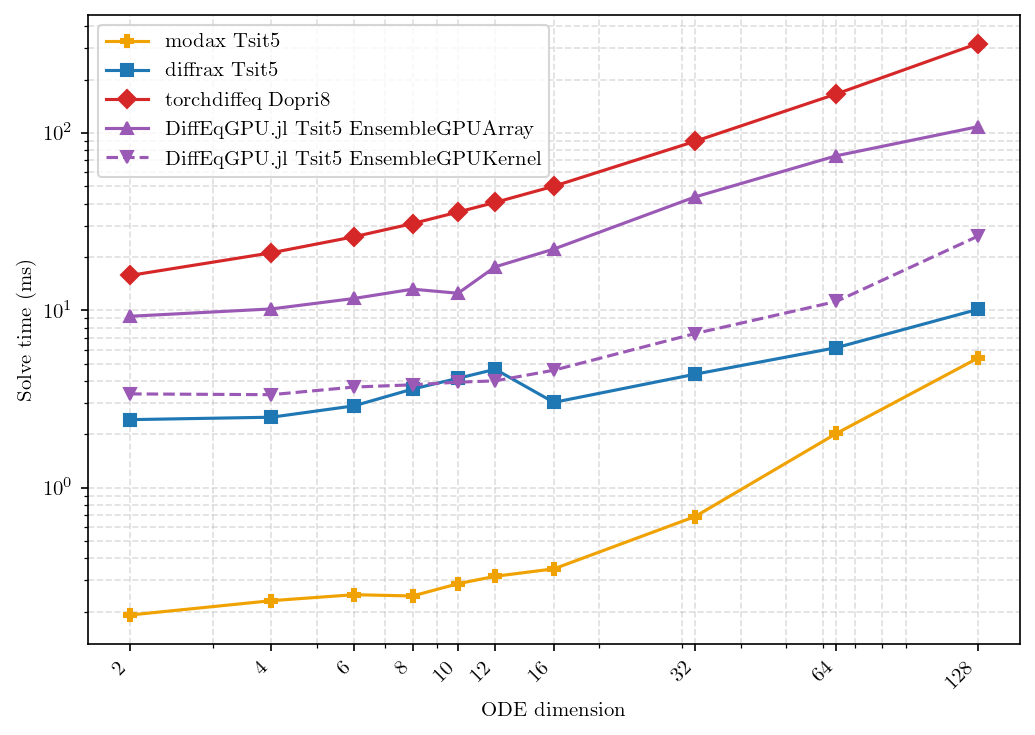}
  \caption{Effect of state dimension $D$ on Tsit5 solver performance for 1000 identical non-stiff van der Pol lattice trajectories. \modax{} is the fastest at every dimension. Its solve time steps up past $D = 16$, where the kernel falls back to storing all stage vectors in global memory. The dip in the \diffrax{} timings near $D = 12$ appears to be a backend shape-specialisation effect rather than a change in numerical work, as the measured step counts remain similar to those for neighbouring dimensions.}
  \label{fig:vdp_tsit5_dim}
\end{figure}

Figure~\ref{fig:vdp_tsit5_dim} shows the effect of dimensionality on the Tsit5 solver. \modax{} scales approximately linearly with $D$ (since $f$ consists of $\mathcal{O}(D)$ arithmetic operations per stage and no matrix factorisations are required), and remains the fastest solver across the entire tested range. It stores all state and stage vectors in faster shared memory at lower dimensions and smaller ensemble sizes, but at higher dimensions falls back to slower, but more abundant, thread-local memory, creating the observed kink between dimensions 16 and 64. We choose a shared-to-local memory transition somewhat smaller than what actually saturates the shared memory budget in order to allow for higher warp occupancy: an SM can hold multiple resident warps of 32 trajectories, and if one warp is stalled waiting for DRAM, the scheduler can switch to another warp that is ready to execute.

\begin{figure}
  \centering
  \includegraphics[width=\columnwidth]{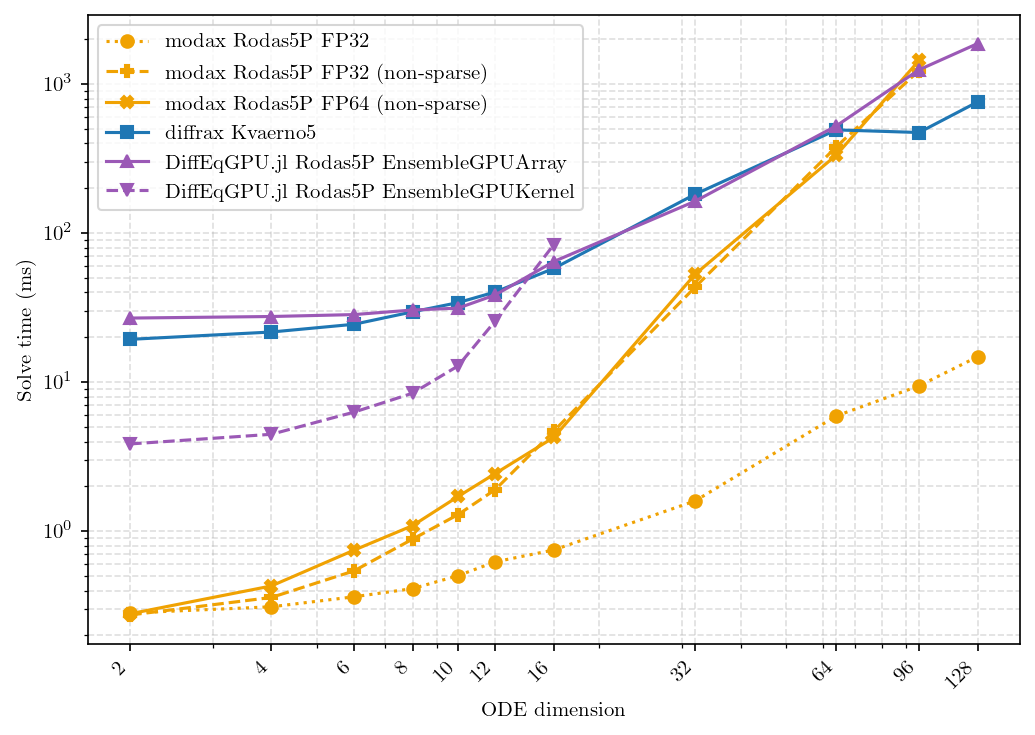}
  \caption{Effect of state dimension $D$ on Rodas5P solver performance for 1000 identical stiff van der Pol lattice trajectories. In its baseline configuration \modax{} is the fastest at every dimension; the non-sparse configurations disable the sparsity analysis. \diffequgpu{}'s EnsembleGPUKernel appears to exceed register/cache limits and fails above $D = 16$.}
  \label{fig:vdp_rodas5_dim}
\end{figure}

Figure~\ref{fig:vdp_rodas5_dim} shows the corresponding analysis for Rodas5P. With sparsity analysis disabled, the implicit solver must store and factorise a $D \times D$ Jacobian matrix, so memory requirements scale as $\mathcal{O}(D^2)$ per trajectory and computation as $\mathcal{O}(D^3)$ per step. The fact that the solver using 32-bit matrix factorisation performs similarly to the 64-bit solver suggests that either the solver is memory-bound rather than compute-bound, or reduction in method order due to the inexact Jacobian may be causing additional steps to be taken \citep{Steinebach2023}. Efficiently exploiting the sparsity of the van der Pol system is therefore crucial to scaling to higher dimensions. The van der Pol lattice has a banded sparsity pattern, so the number of colours required for forward-mode differentiation is constant with respect to dimension, and the cost of computing the Jacobian grows only linearly with $D$. The baseline FP32 \modax{} solver is therefore able to maintain a consistent 100$\times$ speed-up over \diffrax{}'s Kvaerno5 solver across the entire range of dimensions.

\begin{figure}
  \centering
  \includegraphics[width=\columnwidth]{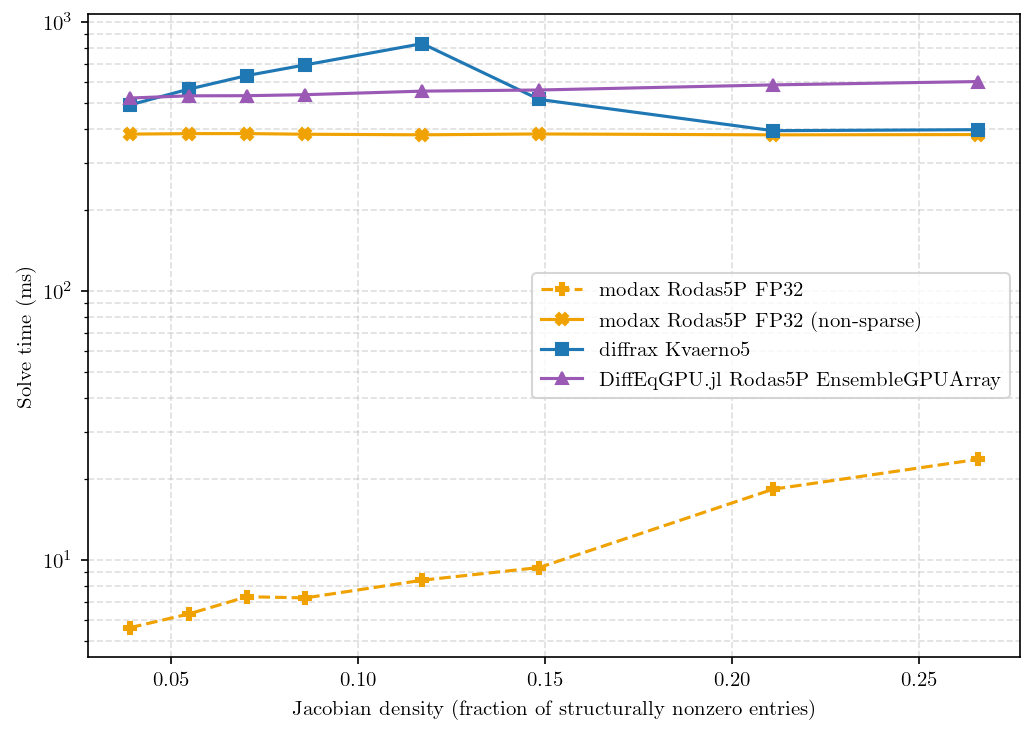}
  \caption{Effect of Jacobian density on Rodas5P solver performance for 1000 identical stiff van der Pol lattice trajectories at fixed dimension $D = 64$. The coupling range of the lattice is varied from nearest-neighbour to all-to-all, changing the number of structurally non-zero Jacobian entries without changing $D$. The non-sparse solvers are insensitive to density, while the sparsity-aware baseline \modax{} solver slows as the pattern fills in. \diffequgpu{}'s EnsembleGPUKernel fails to compile and solve within the benchmark's 180\,s per-point cap at $D = 64$, consistent with Fig.~\ref{fig:vdp_rodas5_dim}.}
  \label{fig:vdp_sparsity}
\end{figure}

The dimensionality sweep above holds the coupling of the lattice fixed at nearest neighbours, so the Jacobian is always banded and its density falls as $1/D$. In Fig.~\ref{fig:vdp_sparsity} we instead fix the dimension at $D = 64$ and vary the coupling range -- the number of places along the ring, on either side, within which oscillators are coupled -- from nearest neighbour to all-to-all, which allows us to investigate how our solvers perform on systems with identical dimensionality but different density. The non-sparse \modax{} solver, and its counterparts from \diffrax{} and \diffequgpu{}, are, as expected, insensitive to density: they always differentiate, store and factorise the full $64 \times 64$ matrix. The baseline sparsity-aware \modax{} solver, on the other hand, slows down as the pattern fills in -- each additional coupled neighbour increases both the number of colour sweeps required to compute the Jacobian and the fill-in of the sparse LU factorisation. At 25 per cent density it is still over 15$\times$ faster than the non-sparse solvers. It is important to note that \modax{}'s ability to exploit sparsity is not confined to banded systems, but extends to any arbitrary pattern in which a substantial fraction of the Jacobian entries are known to be zero.

\begin{figure}
  \centering
  \includegraphics[width=\columnwidth]{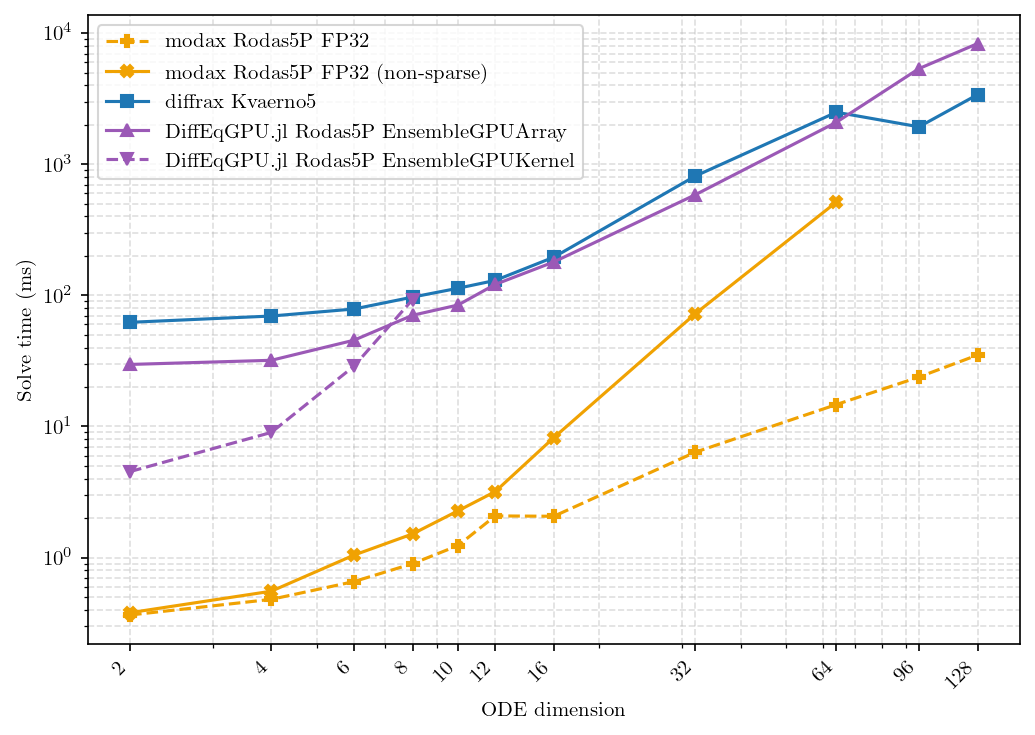}
  \caption{Effect of state dimension $D$ on the cost of computing both the value and gradient of the final state with respect to each trajectory's damping parameter, for 1000 identical stiff van der Pol lattice trajectories. \modax{} and \diffequgpu{} integrate a single forward sensitivity column jointly with the state whilst \diffrax{} uses its default reverse-mode checkpointing adjoint.}
  \label{fig:vdp_gradient}
\end{figure}

Finally, Fig.~\ref{fig:vdp_gradient} benchmarks differentiation through the stiff solve using the forward sensitivity method described in Section~\ref{sec:sensitivity}. At each dimension, we time how long it takes to compute the value of the final state and its gradient with respect to each trajectory's $\mu$ parameter (the scale factor on the damping coefficient). This corresponds to a single sensitivity column ($m = 1$) being integrated alongside the state. The \modax{} solvers are differentiated simply by wrapping the solve in \texttt{jax.value\_and\_grad}, which lowers to a single kernel launch integrating the joint system~\eqref{eq:joint}. \diffrax{} is differentiated in the same way, but in reverse mode through its default recursive checkpointing adjoint. \diffequgpu{} does not provide for differentiation in either of its ensemble backends, so we instead augmented it with forward sensitivity equations by hand. The sparsity-aware baseline \modax{} solver computes the value and gradient at around 2$\times$ the cost of the solve without the gradient (Fig.~\ref{fig:vdp_rodas5_dim}). The additional cost comprises one forward automatic differentiation sweep and one pair of triangular solves per stage for the sensitivity column. It is roughly two orders of magnitude faster than both \diffrax{} and \diffequgpu{}'s EnsembleGPUArray across the entire range of dimensions.

\subsection{Practical examples}
\label{sec:cosmology_examples}

Having benchmarked the solvers on several canonical benchmark systems, we now demonstrate \modax{} on three end-to-end problems drawn from cosmology, each chosen to exercise one of the applications described in Section~\ref{sec:background}: Bayesian parameter estimation, Monte Carlo uncertainty quantification, and the simulation of uncoupled physical systems. All three are provided as runnable examples in the \modax{} repository.

For each example we report the wall-clock cost of the core ensemble solve under three solver backends running on identical right-hand-side code, selectable through a single command-line switch in the example script. The first is the GPU-batched \modax{} solver used by the example. The second is plain \diffrax{} on the same GPU under \texttt{jax.vmap}: the two stiff examples use Kvaerno5, since \diffrax{} provides no Rosenbrock--Wanner method, and the non-stiff example uses Tsit5. The third is \texttt{scipy.integrate.solve\_ivp}, integrating each trajectory independently on the CPU -- LSODA for the stiff examples and RK45 for the non-stiff one -- representative of the integration strategy used by established serial cosmology codes such as ECHO21 \citep{Mittal2026}.

To make the comparisons equivalent we ran the tests on a GPU and CPU that were similar in cost and maximum power consumption. For the GPU, we use the same NVIDIA GeForce RTX 4070 SUPER as in Section~\ref{sec:results} (7,168 CUDA cores with max power draw of 220\,W). For the CPU, we use an AMD Ryzen 9 7950X (16 physical cores / 32 hardware threads with max power draw of 230\,W) and parallelise the SciPy ensemble across all cores with Python's \texttt{multiprocessing} module.

The resulting timings are collected in Table~\ref{tab:cosmology_timing}. They exclude one-time JIT compilation and are the mean of three repeated solves. All three backends run at the same tolerances. For each example we report how well the backends agree on the physical observable at that tolerance.

\subsubsection{Bayesian parameter estimation from primordial abundances}
\label{sec:bbn}

Our first example illustrates the application of \modax{} to parameter estimation: we infer cosmological parameters from primordial light-element abundances by fitting a big bang nucleosynthesis (BBN) model. The forward model we employ is a phenomenological stiff toy network of four species -- neutrons, protons, deuterium, and helium-4 -- which we use both to generate synthetic observations and to fit them. It is intended as a demonstration of the solver on a stiff inference problem, not as a realistic BBN constraint.

Writing the state as the vector of nucleon number fractions $Y = (Y_n, Y_p, Y_d, Y_{^4\mathrm{He}})$ and using $\mathrm{d}x/\mathrm{d}t = Hx$ to change the independent variable from time to $x = Q/T$, where $Q = 1.293\,\mathrm{MeV}$ is the neutron--proton mass difference, the network reads:
\begin{align}
  \frac{\mathrm{d}Y_n}{\mathrm{d}x} &= \frac{\Gamma_{p\to n}\,Y_p - \Gamma_{n\to p}\,Y_n - R_{np}}{Hx}, \label{eq:bbn_n}\\[2pt]
  \frac{\mathrm{d}Y_p}{\mathrm{d}x} &= \frac{\Gamma_{n\to p}\,Y_n - \Gamma_{p\to n}\,Y_p - R_{np}}{Hx}, \label{eq:bbn_p}\\[2pt]
  \frac{\mathrm{d}Y_d}{\mathrm{d}x} &= \frac{R_{np} - 2R_{dd}}{Hx}, \label{eq:bbn_d}\\[2pt]
  \frac{\mathrm{d}Y_{^4\mathrm{He}}}{\mathrm{d}x} &= \frac{R_{dd}}{Hx}. \label{eq:bbn_he}
\end{align}

Neutrons and protons interconvert through the weak rate of \citet{Bernstein1989}, deuterium is driven towards its Saha equilibrium value, and the net $n$--$p$ and $d$--$d$ reaction rates evolve the heavier species, subject to the conservation law $Y_n + Y_p + 2Y_d + 4Y_{^4\mathrm{He}} = 1$. All rates are approximate closed-form analytical functions of $x$ and the cosmological parameters rather than tabulated or interpolated quantities. $H$ is the Hubble expansion rate of the radiation-dominated Friedmann universe through which $N_\mathrm{eff}$ enters the model via the effective number of relativistic degrees of freedom $g_*$.

Precision predictions of the primordial abundances usually require a more comprehensive reaction network. The classic Fortran code \textsc{PArthENoPE} \citep{Pisanti2008PArthENoPE} integrates a network of roughly a hundred reactions, \textsc{PRIMAT} \citep{Pitrou2018PRIMAT} enlarges the network and refines the weak-interaction corrections that set $Y_P$, and the more recent Python packages \textsc{PRyMordial} \citep{Burns2024PRyMordial} and \textsc{LINX} \citep{Giovanetti2025LINX} are designed for fast parameter inference within and beyond the standard model. \textsc{LINX} in particular is written in JAX and integrates its network with \diffrax{} solvers, making it a natural candidate to benefit from the batched implicit solvers presented here. Our toy network omits the tritium and helium-3 channels through which most helium-4 is actually synthesised, so its abundance predictions are far from the observed ones: at the Planck 2018 values $\eta_{10} \approx 6.1$ and $N_\mathrm{eff} = 3.044$ it yields $Y_P = 0.106$ and $D/H = 4.2 \times 10^{-6}$, whereas the standard values are $Y_P \approx 0.245$ \citep{Aver2015} and $D/H \approx 2.5 \times 10^{-5}$ \citep{Cooke2018}. Its posteriors therefore carry no cosmological weight. Nevertheless, the toy network captures the essential stiff dynamics of BBN that arise from the wide separation between the fast weak-interaction rates and the slower nuclear burning, and serves as an effective test case for our GPU-accelerated implicit Rodas5P solver.

We place uniform priors on the baryon-to-photon ratio, $\log_{10}(\eta_{10}) \in [0.5, 1.0]$, and the effective number of neutrino species, $N_\mathrm{eff} \in [2, 4]$, and fit the model to synthetic primordial helium mass fraction $Y_P$ and deuterium ratio $D/H$ observations generated from the toy network itself at the Planck 2018 values, namely the $Y_P = 0.106$ and $D/H = 4.2 \times 10^{-6}$ quoted above, with Gaussian uncertainties of about $4$ and $7$ per cent respectively, comparable to the relative precision of the real measurements. We sample using the nested slice sampling algorithm of \citet{Yallup2026}, a vectorised form of nested sampling \citep{Skilling2006} designed for execution on GPUs and available as part of the \textsc{BlackJAX} package. Each likelihood evaluation requires a forward integration of the BBN reaction network for a proposed parameter vector, and the sampler evaluates its population of proposals under \texttt{jax.vmap}, which the batching rule of the \modax{} solver turns into a single ensemble solve.

\begin{figure*}
  \centering
  \includegraphics[width=\textwidth]{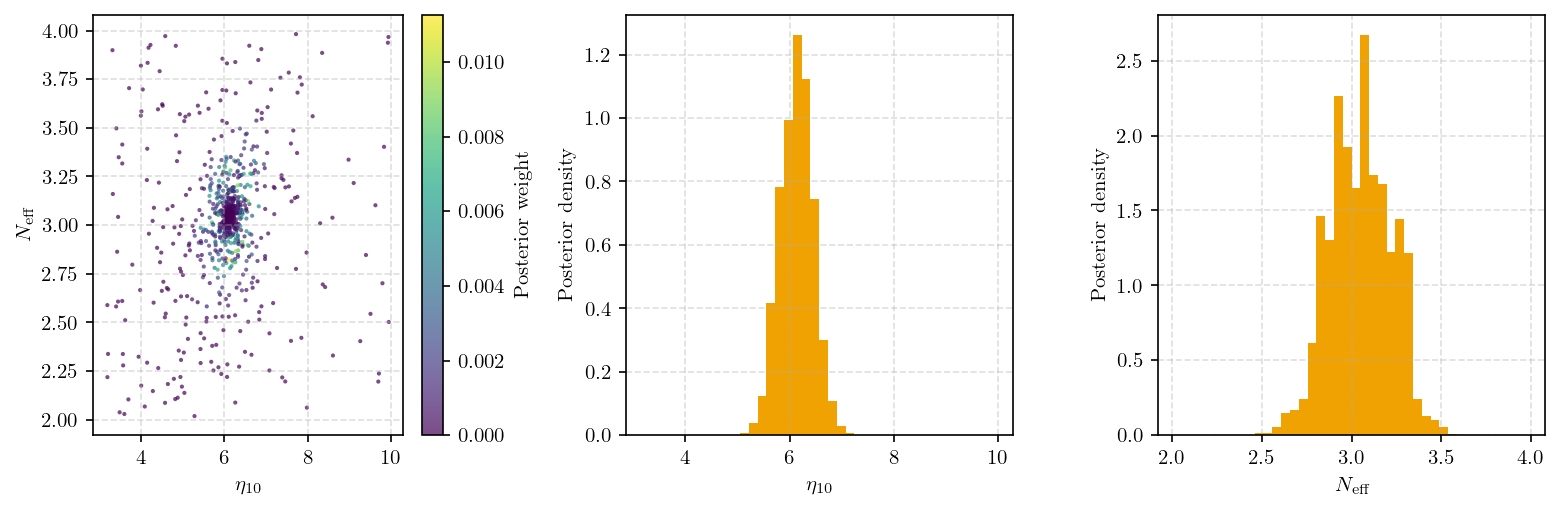}
  \caption{Posterior recovered by nested slice sampling for the phenomenological toy BBN network, fitting synthetic $Y_P$ and $D/H$ observations generated from the same network at $\eta_{10} \approx 6.1$ and $N_\mathrm{eff} = 3.044$. Left: the dead and live points of the run coloured by posterior weight; the lightly weighted points scattered across the prior box are the early, low-likelihood dead points that nested sampling discards on its way to the peak. Centre and right: the weighted marginal posteriors of $\eta_{10}$ and $N_\mathrm{eff}$.}
  \label{fig:bbn_posterior}
\end{figure*}

With $128$ live points the sampler recovers $\eta_{10} = 6.14 \pm 0.31$ and $N_\mathrm{eff} = 3.05 \pm 0.17$, squarely on the calibration values (Fig.~\ref{fig:bbn_posterior}). The example integrates at $\mathrm{rtol} = 10^{-4}$ and $\mathrm{atol} = 10^{-9}$, at which the three backends agree on $Y_P$ at $(\eta_{10}, N_\mathrm{eff}) = (4.11, 3.0)$ to three decimal places ($0.079$). The three backends differ markedly in cost. Integrating a parameter grid of $N = 2000$ stiff universes takes $8.2\,\mathrm{s}$ with \texttt{scipy.integrate.solve\_ivp} when parallelised across all 32 hardware threads ($4090\,\mu\mathrm{s}$ per solve), $1.6\,\mathrm{s}$ with \diffrax{} Kvaerno5 ($802\,\mu\mathrm{s}$ per solve), and just $8\,\mathrm{ms}$ with \modax{} Rodas5P ($4.2\,\mu\mathrm{s}$ per solve): a factor of roughly $970$ over the cost-equivalent parallel CPU baseline and $190$ over \diffrax{}.

\subsubsection{Monte Carlo uncertainty quantification of the global 21\,cm signal}
\label{sec:21cm}

Our second example illustrates the Monte Carlo uncertainty-quantification pattern: we propagate uncertainty in the astrophysics of the first galaxies through to the predicted sky-averaged 21\,cm brightness temperature. We draw uncertain astrophysical parameters -- the star-formation efficiency $\log_{10} f_\star$, the X-ray efficiency $\log_{10} f_X$, and the minimum virial temperature $\log_{10} T_\mathrm{vir}$ -- from their priors and evolve a large ensemble (by default $5 \times 10^4$) of independent intergalactic-medium thermal and ionisation histories, reporting the median and 95 per cent prior envelope of the brightness temperature as a function of observed frequency.

Each trajectory integrates a three-component state -- the log gas kinetic temperature $\log T_K$ and the logit-transformed residual electron fraction $x_e$ and ionised volume-filling fraction $Q_\mathrm{HII}$ -- over the redshift-derived variable $u = \log[(1 + z_\mathrm{initial})/(1 + z)]$. Writing the dynamics in cosmic time $t$, to which the integration variable is related by $\mathrm{d}u = H\,\mathrm{d}t$, the three components obey
\begin{align}
  \frac{\mathrm{d}T_K}{\mathrm{d}t} &= -2H\,T_K + \Gamma_C\,(T_\gamma - T_K) + \epsilon_X + \epsilon_\alpha, \label{eq:21cm_T}\\[2pt]
  \frac{\mathrm{d}x_e}{\mathrm{d}t} &= -\alpha_B(T_K)\,n_H\,x_e^2 + \gamma_X\,(1 - x_e), \label{eq:21cm_xe}\\[2pt]
  \frac{\mathrm{d}Q_\mathrm{HII}}{\mathrm{d}t} &= \Lambda_\mathrm{ion}\,(1 - Q_\mathrm{HII}) - \mathcal{R}\,Q_\mathrm{HII}^2, \label{eq:21cm_Q}
\end{align}
combining, in turn, adiabatic cooling $-2H\,T_K$, Compton coupling to the CMB at temperature $T_\gamma = T_{\gamma,0}(1+z)$, and X-ray and Lyman-$\alpha$ heating in the temperature equation; case-B recombination and X-ray ionisation in the residual free-electron equation; and a stellar ionising-emissivity source balanced against recombinations in the filling-fraction equation \citep{Field1958, Furlanetto2006, Pritchard2012}. The heating terms $\epsilon_X$ and $\epsilon_\alpha$, and likewise the X-ray ionisation rate $\gamma_X$ and ionising emissivity $\Lambda_\mathrm{ion}$, are analytical approximations rather than tabulated or interpolated quantities.

This is a surrogate intended for a batching demonstration, not a precision 21\,cm-signal code. The \modax{} solve uses the per-component error weights of Section~\ref{sec:implementation}, weighting the two logit-transformed fractions at $0.2$ relative to the temperature, together with a proportional--integral step controller; the reference backends run with their default controllers. Because the trajectories are completely independent, this is an ideal showcase for massive batching: $5 \times 10^4$ uncoupled small ODEs are solved in a single call, and the spread of step counts across the prior volume places the ensemble in the divergent regime characterised in Section~\ref{sec:results_robertson}, in which the per-thread \modax{} kernel performs better than batched solvers such as \diffrax{}.

\begin{figure}
  \centering
  \includegraphics[width=\columnwidth]{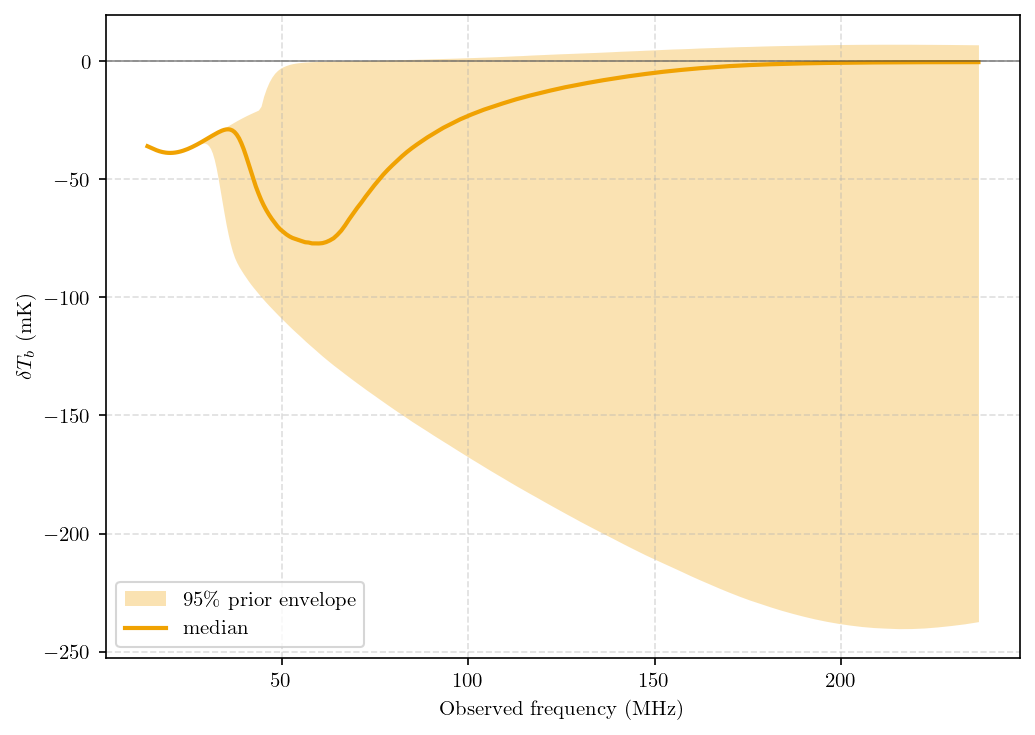}
  \caption{Median (line) and $95$ per cent prior envelope (shaded) of the sky-averaged 21\,cm brightness temperature $\delta T_b$ as a function of observed frequency, from $5 \times 10^4$ intergalactic-medium histories with astrophysical parameters drawn from the priors of Section~\ref{sec:21cm}. The median shows the characteristic absorption trough at $59\,\mathrm{MHz}$ ($z \approx 23$). The envelope widens towards low redshift, corresponding to high frequency, as the uncertain X-ray heating and ionising emissivity take effect.}
  \label{fig:21cm_envelope}
\end{figure}

The ensemble reproduces a median absorption trough of $-77\,\mathrm{mK}$ at $59\,\mathrm{MHz}$ ($z \approx 23$), with a $95$ per cent prior envelope spanning $-122$ to $0\,\mathrm{mK}$ at that frequency (Fig.~\ref{fig:21cm_envelope}); the three backends place the median trough within $0.1\,\mathrm{mK}$ of one another. For a head-to-head ensemble of $N = 20\,000$ histories the parallel CPU solve takes $13.4\,\mathrm{s}$ ($671\,\mu\mathrm{s}$ per solve), \diffrax{} Kvaerno5 takes $1.45\,\mathrm{s}$ ($73\,\mu\mathrm{s}$ per solve), and \modax{} Rodas5P takes $31\,\mathrm{ms}$ ($1.6\,\mu\mathrm{s}$ per solve): a factor of roughly $430$ over the cost-equivalent parallel CPU baseline and $47$ over \diffrax{}. The full $5 \times 10^4$-history Monte Carlo solve of the example completes in $60\,\mathrm{ms}$ on \modax{} against $2.8\,\mathrm{s}$ on \diffrax{}.

\subsubsection{Uncoupled Fourier modes: the primordial power spectrum}
\label{sec:mukhanov_sasaki}

Our final example illustrates an application to physically uncoupled systems, specifically the primordial scalar power spectrum from single-field quadratic inflation following the conventions of \citet{Baumann2009}. First we integrate the homogeneous inflationary background in e-fold time $N = \log a$ to build interpolation tables for the slow-roll parameter $\epsilon(N)$, the comoving Hubble scale $aH(N)$, and the effective mass term $z''/z$ formed from $z(N) = a\,\mathrm{d}\phi/\mathrm{d}N$. Then we solve the Mukhanov--Sasaki perturbation equation \citep{Mukhanov1992},
\begin{equation}
  \frac{\mathrm{d}^2 v_k}{\mathrm{d}N^2}
  = -(1-\epsilon)\frac{\mathrm{d} v_k}{\mathrm{d}N}
  - \left[\left(\frac{k}{aH}\right)^2 - \frac{z''/z}{(aH)^2}\right] v_k,
  \label{eq:mukhanov_sasaki}
\end{equation}
for a large ensemble of comoving wavenumbers $k$. Because $v_k$ is complex while the solver operates on real states, each mode carries the four-component state $y = [\mathrm{Re}\,v_k,\, \mathrm{Im}\,v_k,\, \mathrm{Re}\,v_k',\, \mathrm{Im}\,v_k']$, initialised in the Bunch--Davies vacuum \citep{BunchDavies1978} while deep inside the horizon.

Each $k$-mode is an entirely independent linear ODE, so all modes are batched into a single Tsit5 solve. Crucially, each mode has its own start and stop e-fold -- set by horizon crossing $k/aH = X$ -- so we map every trajectory onto a normalised solver time $s \in [0,1]$ and rescale the right-hand side by $\mathrm{d}N/\mathrm{d}s$ accordingly. This is a clean demonstration of solving an ensemble with per-trajectory integration intervals in a single batched call. Concretely, the background is a quadratic potential $V = \tfrac{1}{2} m^2 \phi^2$ started at $\phi_0 = 17.5$ in reduced Planck units and integrated over $N \in [0, 80]$ e-folds (2000 save points, $\mathrm{rtol}=10^{-9}$, $\mathrm{atol}=10^{-11}$), with the inflaton mass $m$ fixed by normalising the scalar amplitude to $A_s = 2.1\times10^{-9}$ at the pivot scale $k_\star = 0.05\,\mathrm{Mpc}^{-1}$, taken to exit the horizon $55$ e-folds before the end of inflation. The example's default run resolves $41$ comoving wavenumbers spaced geometrically over $k \in [10^{-4}, 1]\,\mathrm{Mpc}^{-1}$, and the benchmark ensemble $N = 4096$ over the same range; each mode is initialised in the Bunch--Davies vacuum deep inside the horizon at $k/aH = 80$ and integrated until it is well outside the horizon at $k/aH = 10^{-3}$, at $\mathrm{rtol} = 10^{-7}$ and $\mathrm{atol} = 10^{-9}$. From the curvature perturbation $\mathcal{R}_k = v_k/z$ we extract the dimensionless power spectrum $\mathcal{P}_\mathcal{R}(k) = (k^3/2\pi^2)\,|v_k/z|^2$, and from it the scalar amplitude $A_s$ and spectral index $n_s$, which we compare against the slow-roll estimates. This is precisely the same kind of structure as the mode-by-mode linearised Einstein--Boltzmann solve used in CMB Boltzmann codes such as \textsc{camb} \citep{Lewis2000} and \textsc{class} \citep{Lesgourgues2011}, and a natural target for the per-component error weighting discussed in Section~\ref{sec:implementation}.

The numerical spectrum yields a scalar amplitude $A_s = 2.067\times10^{-9}$ and spectral index $n_s = 0.9633$, within $0.3$ per cent and $0.0006$ of the slow-roll estimates $A_s = 2.061\times10^{-9}$ and $n_s = 0.9640$ respectively, and the pivot amplitude agrees across all three backends to five significant figures. Because the mode equation is non-stiff and strongly oscillatory we compare against explicit solvers: \modax{} Tsit5, \diffrax{} Tsit5, and \texttt{scipy.integrate.solve\_ivp} RK45. Resolving $4096$ modes at $\mathrm{rtol}=10^{-7}$ takes the parallel CPUs $53\,\mathrm{s}$ ($13\,000\,\mu\mathrm{s}$ per solve), whilst \diffrax{} Tsit5 takes $106\,\mathrm{ms}$ ($26\,\mu\mathrm{s}$ per solve) and \modax{} Tsit5 takes $9\,\mathrm{ms}$ ($2.3\,\mu\mathrm{s}$ per solve): a factor of roughly $5700$ over the cost-equivalent parallel CPU baseline and $11$ over \diffrax{}.

While the power spectrum requires only a single mode function per wavenumber, the same Mukhanov--Sasaki solutions are the computational bottleneck when it comes to predicting the primordial \emph{bispectrum} -- the three-point correlator of the curvature perturbation that is the leading probe of primordial non-Gaussianity and a sharp discriminant between inflationary models \citep{Fergusson2009}. For general models, beyond the slow-roll regime or with features in the potential, the bispectrum has no closed form and must be evaluated numerically, with an integrand built from the mode functions $v_k(N)$ and their derivatives. Because the bispectrum is a function on the three-dimensional space of wavenumber triangles $(k_1, k_2, k_3)$, and modal-expansion estimators sample it on a dense grid of such configurations \citep{Fergusson2010}, a single shape prediction can require the mode functions to be evaluated for thousands of distinct wavenumbers, with the whole computation repeated across the model parameter space during inference or forecasting. This is a massive ensemble of independent linear ODEs of exactly the form solved here. Our faster ensemble solvers can therefore help address this recognised computational bottleneck in studies of primordial non-Gaussianity.

\begin{table}
  \centering
  \caption{Per-solve wall-clock cost of the core ensemble integration in each worked cosmological example, comparing \texttt{scipy.integrate.solve\_ivp} parallelised across all 32 hardware threads of an AMD Ryzen~9 7950X CPU, plain \diffrax{} on an NVIDIA GeForce RTX~4070 SUPER GPU, and \modax{} on the same GPU, all using identically implemented ODE functions. The CPU and GPU have broadly comparable retail price and operating power so this is a like-for-like comparison. The stiff examples (BBN, 21\,cm) use implicit methods (\modax{} Rodas5P, \diffrax{} Kvaerno5, SciPy LSODA) whilst the non-stiff Mukhanov--Sasaki example uses explicit methods (Tsit5, Tsit5, SciPy RK45). Timings exclude JIT compilation and are the mean of three repeated runs after a warm-up. Costs are per trajectory in microseconds, for ensembles of $2000$ (BBN), $20\,000$ (21\,cm) and $4096$ (Mukhanov--Sasaki) trajectories. The final two columns are the \modax{} speed-up over the cost-equivalent parallel CPU baseline and over \diffrax{}.}
  \label{tab:cosmology_timing}
  \footnotesize
  \setlength{\tabcolsep}{2pt}
  \begin{tabular*}{\columnwidth}{@{\extracolsep{\fill}}lccccc@{}}
    \toprule
    Example & SciPy & \diffrax{} & \modax{} & \multicolumn{2}{c}{\modax{} speed-up} \\
    & (CPU) & (GPU) & (GPU) & vs CPU & vs \diffrax{} \\
    \midrule
    BBN (4D, stiff)          & 4090    & 802 & 4.2 & $970\times$  & $190\times$ \\
    21\,cm (3D, stiff)       & 671     & 73  & 1.6 & $430\times$  & $47\times$  \\
    Mukhanov (4D, non-stiff) & 13\,000 & 26  & 2.3 & $5700\times$ & $11\times$  \\
    \bottomrule
  \end{tabular*}
\end{table}

\section{Conclusions}
\label{sec:conclusions}

We have presented \modax{}, a GPU-accelerated ODE solver library in Python optimised for solving large ensembles of low-dimensional ODEs that differ only in parameters or initial conditions. It implements a Tsit5 explicit and a Rodas5P implicit solver using thread-level CUDA programming to allow for kernel fusion and near-independent evolution of ODE trajectories. Our solvers are between 1 and 3 orders of magnitude faster than their counterparts in \diffrax{} and \diffequgpu{}, as benchmarked on a range of canonical test problems and three end-to-end cosmological examples: Bayesian parameter estimation from primordial abundances, Monte Carlo uncertainty quantification of the global 21\,cm signal, and the computation of uncoupled Fourier modes in the primordial power spectrum. Their design reflects the importance of parallelism across independent scientific calculations, with automatic differentiation providing an additional capability alongside efficient ensemble evaluation.

For solving stiff systems, \emph{linearly implicit} Rosenbrock--Wanner methods such as our implementation of Rodas5P stand out as being particularly well suited to GPUs. By baking the Jacobian directly into the method's algebraic structure, a Rosenbrock--Wanner step replaces the non-linear stage equations of a Runge--Kutta method with a fixed sequence of linear solves that share a single per-step factorisation, eliminating Newton iteration and with it the thread divergence of data-dependent iteration counts. Its order conditions are constructed so that at least second order is retained even when the Jacobian used is only an approximation. This opens the door to performing computationally demanding factorisations in single rather than double precision and thereby exploiting the far higher single-precision throughput of modern GPUs. Further exploiting sparsity patterns of the Jacobian allows our implementation to scale to higher-dimensional systems than would otherwise be possible given the limited fast memory and bandwidth of GPUs.

In the future, we envisage extending our CUDA-compiled batched ensemble machinery to stochastic differential equations, with applications spanning stochastic chemical kinetics, Langevin dynamics and the score-based diffusion models that dominate modern generative modelling, as well as neural ODEs, whose training and inference amount to repeated integration of large ensembles of parameterised ODEs, with applications to low-dimensional continuous normalising flows and irregularly sampled time-series modelling. We also plan on building a high-order implicit--explicit Rosenbrock--Wanner solver, which would combine the structural advantages of IMEX methods with the iteration-free Wanner framework. Such methods are particularly useful for problems with additive stiffness, allowing for efficient integration by treating the stiff component implicitly and the non-stiff component explicitly. However, existing IMEX methods require iterative non-linear solves, which, as discussed, can introduce thread divergence and reduce performance on GPUs. By developing a Rosenbrock--Wanner IMEX solver, we aim to retain the benefits of separating the stiff and non-stiff parts of a system while eliminating the risk of thread divergence on GPUs.

Looking towards the future of GPU scientific computing, it is worth noting that newer hardware appears to be prioritising lower-precision arithmetic throughput over higher precision, driven by the demands of machine learning workloads which often operate at lower precision. The latest NVIDIA Blackwell Ultra (B300) GPUs retain only $1/30$th of the double-precision throughput of the preceding Blackwell B200 and Hopper H100 generations \citep{NVIDIAH100, NVIDIAHGX2026}. In light of this, it will become increasingly important to consider the role that single-precision and mixed-precision arithmetic can play in scientific computing, and to design algorithms that can take advantage of lower precision or ``tensor-core'' matrix-multiplication compute without sacrificing accuracy.

\section*{Acknowledgements}

The authors thank the open-source developers of \textsc{JAX}, \textsc{Numba-CUDA-MLIR}, \diffrax, \torchdiffeq{} and \diffequgpu{}, whose work provided both a foundational reference and points of comparison for this study. OH thanks Tobias Heinzle for valuable discussions. LB thanks James Alvey for his mentorship and feedback on early drafts of this work.

\section*{Data Availability}

All benchmark code, solver implementations, and data required to reproduce the results presented in this paper are available at \url{https://github.com/lawrenceberry/modax}. The repository includes the full \modax{} library, benchmark scripts, and a test suite verifying solver correctness against analytic solutions.


\bibliographystyle{rasti}
\bibliography{paper}


\bsp
\label{lastpage}
\end{document}